\documentclass[11pt]{article}

\usepackage{fullpage}
\usepackage{amsmath}
\usepackage{amssymb}
\usepackage{setspace}
\usepackage{bbm}
\usepackage{dsfont}
\usepackage{graphics}
\usepackage{cite}
\usepackage{hyperref}

\hypersetup{
    bookmarks=true,%
    colorlinks,%
    citecolor=blue,%
    filecolor=blue,%
    linkcolor=blue,%
    urlcolor=blue
}

\usepackage[font=footnotesize,labelfont=bf,justification=centerlast,width=.94\textwidth]{caption}
\usepackage{color}

\usepackage[normalem]{ulem}
\usepackage{epsfig}

\newcommand{\be}{\begin{equation}}
\newcommand{\ee}{\end{equation}}
\newcommand{\bes}{\begin{equation*}}
\newcommand{\ees}{\end{equation*}}
\newcommand{\bea}{\begin{eqnarray}}
\newcommand{\eea}{\end{eqnarray}}
\newcommand{\beas}{\begin{eqnarray*}}
\newcommand{\eeas}{\end{eqnarray*}}

\newcommand{\p}{\partial}

\newcommand{\bmat}{\begin{bmatrix}}
\newcommand{\emat}{\end{bmatrix}}

\newcommand{\slt}{SL(2,\RR)}

\newcommand{\sln}{SL(N,\RR)}

\newcommand{\RR}{\mathbb{R}}

\def\Tr{{\rm Tr}}
\def\le{\left}
\def\ri{\right}
\def\ep{{\epsilon}}
\newcommand\sO{{\ensuremath{{\mathcal O}}}}

\def\Tr{{\rm Tr}}
\def\le{\left}
\def\ri{\right}

\newcommand{\R}{\mathbb{R}}

\def\le{\left}
\def\ri{\right}
\def\ha{{1\over 2}}

\def\al{{\alpha}}

\def\Tr{{\rm Tr}}

\def\ep{{\epsilon}}

\newcommand\lam{\lambda}
\newcommand\Lam{\Lambda}

\def\lam{{\lambda}}

\def\eeq{\end{equation}}

\def\Tr{\mathop{\rm Tr}}

\newcommand\sL{{\ensuremath{{\mathcal L}}}}

\newcommand\sU{{\ensuremath{{\mathcal U}}}}
\newcommand\sW{{\ensuremath{{\mathcal W}}}}

\newcommand\sR{{\ensuremath{{\mathcal R}}}}

\newcommand\bw{{\bar w}}

\newcommand\ba{{\overline a}}

\newcommand\bB{{\overline{B}}}
\newcommand\bA{\bar{A}}
\newcommand\bb{\bar{b}}
\newcommand\bV{\bar{V}}

\begin{document}
\numberwithin{equation}{section}
{
\begin{titlepage}
\begin{center}

\hfill \\
\hfill \\
\vskip 0.6in

{\Large \bf Eternal Higher Spin Black Holes: a Thermofield Interpretation}\\

\vskip 0.4in

{\large  Alejandra Castro, Nabil Iqbal, and Eva Llabr\'es}\\

\vskip 0.3in

{\it Institute for Theoretical Physics, University of Amsterdam,
Science Park 904, Postbus 94485, 1090 GL Amsterdam, The Netherlands} \vskip .5mm

\end{center}

\vskip 0.45in

\begin{center} {\bf ABSTRACT } \end{center}

We study Lorentzian eternal black holes in the Chern-Simons sector of AdS$_3$ higher spin gravity. We probe such black holes using bulk Wilson lines and motivate new regularity conditions that must be obeyed by the bulk connections in order for the geometry to be consistent with an interpretation as a thermofield state in the dual CFT$_2$. We demonstrate that any higher spin black hole may be placed in a gauge that satisfies these conditions: this is the Chern-Simons analogue of the construction of Kruskal coordinates that permit passage through the black hole horizon. We also argue that the Wilson line provides a higher-spin notion of causality in higher spin gravity that can be used to associate a Penrose diagram with the black hole. We  present some applications of the formalism, including a study of the time-dependent entanglement entropy arising from the higher spin black hole interior and evidence for an emergent AdS$_2$ region in the extremal limit. 


%

\end{titlepage}
}

\newpage

\tableofcontents
\section{Introduction}

Higher spin theories of gravity provide toy models where one can examine ideas of stringy geometry in a controlled setting. In addition to the usual spin-$2$ graviton, such theories contain other, higher spin degrees of freedom that mix nontrivially with the graviton under a large set of gauge redundancies. In three dimensions one can consider theories with only a finite number of higher spin fields, including all spins starting from $2$ up to a fixed highest spin $N$: the relevant bulk description is given by Chern-Simons theory with gauge group $\sln \times \sln$, generalizing the usual presentation of AdS$_3$ gravity as an $\slt \times \slt$ Chern-Simons theory. 

There has been extensive study of black hole solutions in such theories, starting from the work of \cite{Gutperle:2011kf}. Much of the subsequent literature deals with static properties and thermodynamics, and so can largely be thought of as studies of the black hole in its Euclidean section. In this work we study instead the Lorentzian structure of eternal higher spin black holes. In particular, as we review below, it is well-understood in AdS/CFT that an eternal black hole is dual to the thermo-field state in a doubled tensor product of the dual field theory Hilbert space. In what follows, we will discuss the interpretation of eternal higher spin black holes from this point of view. 

In particular, the standard identification of the two-sided black hole with the thermofield state is tied to the causal structure of an eternal black hole. The fact that the two copies of the CFT are decoupled but entangled is roughly dual to the fact that the two boundaries of the eternal black hole are connected -- but not causally so -- by an Einstein-Rosen bridge. To fully flesh out this interpretation in the higher spin case, it would be helpful to give an operational meaning to the ``causal structure'' of an eternal higher spin black hole.  This is a nontrivial endeavour: in higher spin theories, conventional notions of geometry are not even gauge-invariant, and we will require different tools to organize our thinking. 

These theories do not admit a conventional geometric understanding; however they do admit interesting higher-spin-invariant probes. In this paper we will consider the Wilson line operator constructed in \cite{Ammon:2013hba,deBoer:2013vca}. As we review below, this object should be thought of as the higher-spin-invariant generalization of the worldline of a massive particle moving in the bulk, carrying well-defined charges under the higher-spin symmetries. In the simplest case, when it is charged only under the spin$-2$ field -- and thus has a mass but no other charges -- its action in the bulk may thus be thought of as the higher-spin analogue of a bulk proper distance. Furthermore, if the endpoints of this Wilson line are taken to intersect the AdS boundary, it computes both the boundary two-point function of a CFT operator with the specified charges, or (by appropriate choices of these charges) a CFT entanglement entropy \cite{deBoer:2014sna}. 

These Wilson lines then provide us with a sensitive probe of bulk higher spin geometries. Interestingly, we find that the study of Wilson lines on the eternal black hole background requires a refined understanding of regularity properties on the bulk gauge connections. One of our main results is the description of a particular bulk gauge choice -- which we call {\it Kruskal gauge} -- that is in many ways the Chern-Simons analogue of the Kruskal choice of coordinates that permit passage through the event horizon to the full maximally extended spacetime. This gauge choice simply amounts to demanding that the connections be {\it smooth} when evaluated at the Euclidean origin: while this may sound like a very benign condition, it involves an interplay between the bulk radial coordinate and Euclidean time, and so is novel from the point of view of Chern-Simons theory. In particular, it is stronger than the familiar ``holonomy conditions'' of Euclidean regularity that are normally used to define black hole connections: however, given a black hole that satisfies the holonomy condition, there is an algorithm that can be followed to place it into Kruskal gauge. Some recent work that also implements this stronger notion of regularity is in \cite{Banados:2016nkb}.

With an understanding of this bulk gauge choice we then proceed to study the properties of eternal higher spin black holes. We present computations in several gauges to illustrate potential pitfalls, and verify that in Kruskal gauge, all correlators behave as expected for a thermofield state. We also study some of the resulting physics: in particular, we demonstrate that the interior of a two-sided eternal black hole ``grows'' with time (as measured by the action of a bulk Wilson line). We also highlight some interesting features of purely one-sided correlators, studying in particular the behavior of the extremal limit and providing evidence for the emergence of an infrared AdS$_2$.  

Some other recent work involving bulk $U(1)$ Wilson lines that connect the two sides of an eternal black hole includes \cite{Engelhardt:2015fwa,Harlow:2015lma,Guica:2015zpf}. Our viewpoint here is somewhat different from that taken in those works, where a distinction is drawn between the Wilson line operator (which is constructed from the bulk $U(1)$ gauge fields) and the existence of dynamical charged matter in the bulk. However, when 3d gravity is studied in the Chern-Simons formulation, it appears to be impossible to make such a distinction, precisely because there is no simple way to couple Chern-Simons gravity to propagating matter. Our Wilson line should be thought of as providing a geometric optics approximation to the correlation functions of (putative) matter in the bulk, and the interplay of such a Wilson line with {\it actual} dynamical matter is an important topic for future exploration.

The organization of this paper is as follows. We begin in Section \ref{sec:ehsbh} with a review of the usual definition of Euclidean black holes in Chern-Simons theory. In Section \ref{sec:defBH} we motivate the more refined notion of regularity adequate for Lorentzian eternal black holes, defining two forms of the Kruskal gauge mentioned above and explaining their relation. In Section \ref{sec:btz} we apply this formalism to the familiar BTZ black hole and discuss the maximally extended spacetime in the Chern-Simons formalism. In Section \ref{sec:hsbh} we turn finally to the higher spin black hole, where we present computations in several gauges that have appeared in the literature previously as well as in Kruskal gauge. In Section \ref{sec:appl} we discuss some simple applications, including a determination of the entanglement velocity chracterizing the speed of entanglement growth for the higher-spin black hole. We conclude in Section \ref{sec:disc} with a brief discussion and some directions for future research.  


\section{Euclidean higher spin black holes: a review}\label{sec:ehsbh}

In this section we first review the properties of black holes in AdS$_3$ as currently understood in the Chern-Simons formulation of gravity. For a complete discussion and list of references see  \cite{Gutperle:2011kf,Ammon:2011nk,Ammon:2012wc,deBoer:2013gz,Bunster:2014mua,deBoer:2014fra}. 

The black holes that we will study are classical solutions to Chern-Simons theory with a given gauge group. More concretely, the Chern-Simons action is
\begin{equation}
I_{\rm CS} = \frac{ik_{cs}}{4\pi}\int_{M}\text{Tr}\Bigl[{\rm CS}(A)-{\rm CS}(\bar{A})\Bigr]\,, \label{csac}
\end{equation}
 where $A$ and $\bar{A}$ are valued in the same algebra, and 
\begin{equation}
{\rm CS}(A) = A\wedge dA + \frac{2}{3}A\wedge A\wedge A~.
\end{equation}
Our general arguments and results will not be very sensitive to the choice of gauge group, but for the sake of simplicity our explicit computations will involve connections valued in either the Lie algebra $sl(2)$ (in which case we are discussing standard spin-$2$ gravity on AdS$_3$) or $sl(3)$ (in which case we are discussing the simplest theory of higher spin gravity, including a single spin-$3$ field). Gauge transformations $\Lam_{L,R}(x) \in sl(N)$ act as
\be
A \to \Lam_L (A + d) \Lam_L^{-1} \qquad \bA \to \Lam_R^{-1} (\bA + d) \Lam_R \label{bulkgauge}
\ee
In conventional $sl(2)$ gravity, Lorentz transformations form the subgroup with $\Lam_L = \Lam_R^{-1}$ , which rotate the vielbein but leave the metric invariant. 

The equations of motion following from \eqref{csac} simply force both $A$ and $\bA$ to be flat. The standard way to parametrize these flat connections is by gauging away the radial dependence, i.e. 
\begin{equation}\label{eq:aba}
A= b(r)^{-1}\le(a(x^+,x^-) + d\,\ri)b(r)\,,\qquad \bar{A} = b(r)\le(\bar{a}(x^+,x^-) + d\,\ri)b(r)^{-1}\,.
\end{equation}
Here $r$ is the holographic radial direction, and $x^\pm=t\pm \phi$ are the boundary coordinates. In Lorentzian signature we will consider solutions with $\RR \times D_2$ topology; the compact direction on $D_2$ is described by $\phi\sim\phi+2\pi$. In Euclidean signature we will analytically continue $x^\pm$ to complex coordinates $(z,\bar z)$ via $t = i\tau$, and the topology of the bulk is now a solid torus with $z\sim z + 2\pi \sim z + i\beta $. Here $\beta$ is the inverse temperature.\footnote{Throughout this work we will only consider static (non-rotating) solutions, which makes the complex structure of the torus purely imaginary.} $b(r)$ is a radial function that is normally taken to be $e^{r L_0}$: while its precise role in the interior of the geometry is somewhat obscure, its form as $r \to \infty$ is important for the connections to satisfy asymptotically AdS boundary conditions. This will play an important role in what follows. 
 
The connections $a(x^+,x^-) $ and $\bar{a}(x^+,x^-)$ contain the information that characterizes the state in the dual CFT.  In the absence of sources there is systematic procedure to label them: a suitable set of boundary conditions on the connections  results in $\sW$-algebras as asymptotic symmetries \cite{deBoer:1998ip,Henneaux:2010xg,Campoleoni:2010zq,Gaberdiel:2011wb,Campoleoni:2011hg}. These are commonly known as  Drinfeld-Sokolov boundary conditions. To be concrete, for $sl(N)\times sl(N)$ the connections take the form
 \be\label{eq:a12}
 a_z = L_1 + \sum_{s=2}^{N} J_{(s)}(z) W^{(s)}_{-s+1} ~,\quad  \bar a_{\bar z} = L_{-1} + \sum_{s=2}^{N} \bar J_{(s)}(\bar z) W^{(s)}_{s-1} ~,
 \ee
while $a_{\bar{z}} = \bar{a}_{ z}=0$. Here   $\{L_0,L_{\pm1}\}$ are the generators of the $sl(2,\mathbb{R})$ subalgebra in $sl(N)$, and $W^{(s)}_j$ are the spin-$s$ generators with $j=-(s-1),...(s-1)$. $J_{(s)}(z)$ are dimension-$s$ currents whose algebra is  $\sW_N$, and same for the barred sector. 

\begin{figure}
\begin{center}
\includegraphics[width=0.5\textwidth,page=1]{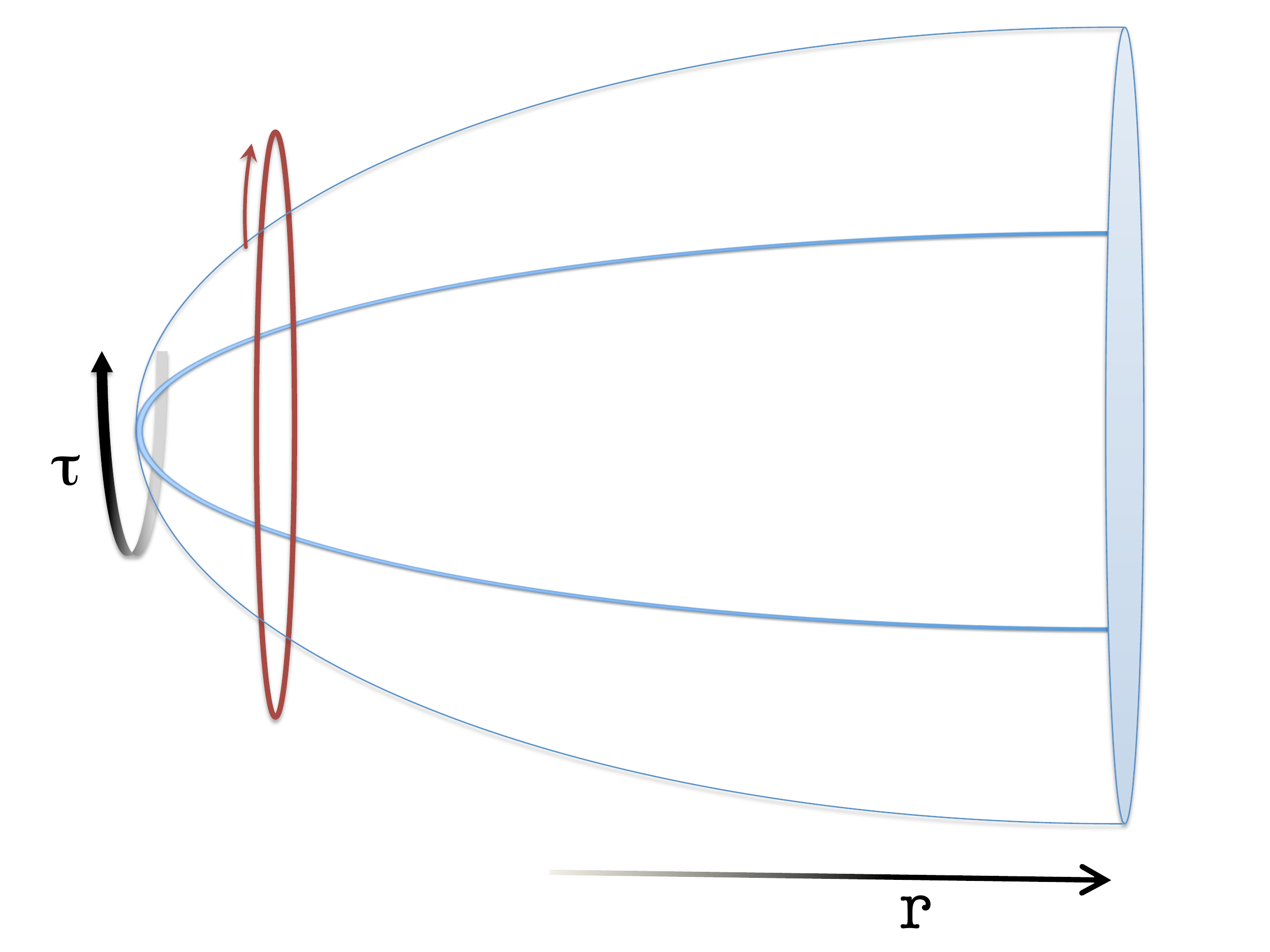}
\end{center}
\caption{Topology of the Euclidean higher spin black hole for a static solution, where the compact direction is Euclidean time $t=i\tau$. The red curve depicts the cycle along which the smoothness condition \eqref{smoothness} is imposed, and it is independent of the radial position. In Euclidean signature, the geometry ends at a finite value of $r$: in a metric-like formulation of gravity this end point would be the horizon.}\label{fig:euc}
\end{figure}

We are interested in stationary black hole solutions, hence $(a,\bar a)$ are constant flat connections that contain both charges and sources. More importantly, the feature that distinguishes black holes  
from other solutions is a smoothness condition. In a metric formulation of gravity, the Euclidean section of a black hole has the property that the compact Euclidean time direction smoothly shrinks to zero size at the horizon of the black hole, resulting in a smooth cigar-like geometry as in Figure \ref{fig:euc}. In the Chern-Simons formulation of gravity, this property is normally thought to generalize to the idea that a black hole is a flat gauge connection defined on a solid torus, where the holonomy along the thermal cycle of the torus belongs to the  center of the group, i.e.
\begin{equation}\label{smoothness}
\mathcal{P}\exp\left(\oint_{\mathcal{C}_E} a\right) \cong e^{\beta a_{\tau}} \cong e^{2\pi i L_0} \,,
\end{equation}
 and similarly in the barred sector; here $L_0$ denotes the Cartan element of $sl(2)$,\footnote{Depending on the gauge group, the choice of center in the rhs of \eqref{smoothness} is not unique \cite{Castro:2011fm}. The choice used here has the feature that it is smoothly connected to the BTZ solution. The interpretations of other choices are discussed in \cite{Hijano:2014sqa,deBoer:2014sna}.} and $\mathcal{C}_E$ is the thermal cycle $z\sim z + i\beta$ which is contractible in the bulk.  

In addition to the smoothness condition, one needs to specify how charges and sources are incorporated in the connections $(a,\bar a)$. From the CFT perspective, it is natural to capture the currents in  $a_{z}$ and the sources in $a_{\bar{z}}$, and vice-versa for $\bar a$ \cite{Gutperle:2011kf}. From the gravitational perspective, the canonical prescription is to encode in $(a_\phi,\bar a_\phi)$ the currents \cite{Banados:2012ue,Perez:2012cf,Compere:2013nba,Henneaux:2013dra}. These two choices, $a_{z}$ versus $a_{\phi}$, amount for different partition functions as shown in \cite{deBoer:2014fra}: the $a_z$ prescription, denoted  {\it holomorphic} black hole, corresponds to a Lagrangian deformation of the theory; the $a_\phi$ prescription,  denoted {\it canonical} black hole, corresponds to a Hamiltonian deformation. It is important to make a distinction between these two, since the Legendre transformation that connects these two prescriptions is non-trivial.

To illustrate these two choices, let us consider black holes in $SL(3)\times SL(3)$ Chern-Simons theory. In this case we define:\footnote{Note that the equations of motion, flatness condition, simply imposes that $[a_+,a_-]=0=[\bar a_-,\bar a_+]$ as can be checked explicit for \eqref{bhconn}.}
 \bea\label{bhconn}
 a_+&=&L_1-\frac{2 \pi  \mathcal{L}}{k}L_{-1}-\frac{\pi \mathcal{W}}{2 k} W_{-2}\,,\nonumber\\\nonumber
 a_-&=&\mu \left(W_2+\frac{4 \pi  \mathcal{W}}{k}L_{-1}+\left(\frac{2 \pi  \mathcal{L}}{k}\right)^2 W_{-2}-\frac{4 \pi  \mathcal{L}}{k}W_0\right)\,,\\ 
 \bar{a}_-&=&-\left(L_{-1}-\frac{2 \pi  \mathcal{L}}{k}L_1+\frac{\pi  \mathcal{W}}{2 k}W_2\right)\,,\\\nonumber 
 \bar{a}_+&=&\mu  \left(W_{-2}-\frac{4 \pi  \mathcal{W}}{k}L_1+\left(\frac{2 \pi  \mathcal{L}}{k}\right)^2 W_2-\frac{4 \pi\mathcal{L}}{k}W_0\right).
 \eea
For simplicity we have turned off rotation, i.e. ${\cal L}={\cal \bar L}$ and ${\cal W}=-{\cal \bar W}$. The interpretation of these connections as thermal states depends on the boundary conditions used to define the classical phase space. The  holomorphic black hole is given by the following connections
 \be\label{hol}
 a_h=a_+dx^+ +a_-dx^-\,,\qquad \bar a_h=\bar a_+dx^+ +\bar a_-dx^-\,,
 \ee
 In this notation the components $(a_+, \bar a_-)$ contain the information of the charges of the system: $({\cal L},{\cal W})$ are the zero modes of the stress tensor and  dimension-3 current of the ${\cal W}_3$ asymptotic symmetry group that organizes the states in this theory. $(\beta,\mu)$  are their respective sources which are fixed by the smoothness condition \eqref{smoothness}. 
 The second prescription, i.e.  the canonical black hole, is given by
 \be\label{can}
 a_c=a_+d\phi +(a_+ + a_-)dt,\qquad \bar a_c=-\bar a_-d\phi +(\bar a_+ +\bar a_-)dt\,. 
 \ee
 For this prescription, again $({\cal L},{\cal W})$ are the zero modes of the currents in ${\cal W}_3$.  The quantitative difference between the holomorphic and canonical definitions lies in the spatial components of the connection; both $a_c$ and $a_h$ have the same time component. The smothness condition \eqref{smoothness} enforces relations between the parameters $\mathcal{L}$, $\mathcal{W}$, $\mu$, and $\beta$. Following \cite{Gutperle:2011kf,Ammon:2011nk}, these constraints can be solved in terms of dimensionless parameter $C\geq3$:
  \be\label{constraintsC}
 \mathcal{W}=\frac{4 (C-1) \mathcal{L}}{C^{3/2}}\sqrt{\frac{2 \pi  \mathcal{L}}{k}}\,,\qquad \mu =\frac{3 \sqrt{C}}{4 (2 C-3)}\sqrt{\frac{k}{2 \pi  \mathcal{L}}}\,,\qquad \frac{\mu}{\beta}=\frac{3}{4\pi}\frac{(C-3)\sqrt{4C-3}}{(3-2C)^2}\,.
 \ee
 The limit $C\rightarrow \infty$ makes the higher spin charges vanish, and we recover the BTZ case;  $C=3$ and $\mu$ fixed corresponds to a zero temperature solution which defines an extremal higher spin black hole \cite{Gutperle:2011kf,Banados:2015tft}. Imposing \eqref{constraintsC}, the eigenvalues of $a_t$ are $\lambda_t= 2\pi L_0/\beta$, for both holomorphic and canonical. In the following sections we will measure $\cal L$ and $\cal W$ in units of $k$; the explicit $k$ dependence will be restored when needed. 
 

 The smoothness condition \eqref{smoothness} is a robust and successful definition of Euclidean black holes. It reproduces in an elegant manner many properties that we expect from a thermal state in the dual CFT$_2$. This definition has also unveiled novel properties of systems in the grand canonical ensemble of ${\cal W}_N$, such as microscopic features of the entropy \cite{Kraus:2011ds,Gaberdiel:2012yb,Compere:2013nba}, ensemble properties \cite{deBoer:2013gz,deBoer:2014fra} and novel phase diagrams \cite{David:2012iu}, and it inspires new observables related to entanglement entropy \cite{deBoer:2013vca,Ammon:2013hba,Hijano:2014sqa}. 
 
It is important to emphasize at this point that the smoothness conditions, the resulting black hole thermodynamics, and the derivation of Ward identities (which identify currents and sources)  are independent of $b(r)$: this could be attributed to the topological nature of the Chern-Simons theory.  As a consequence, these observables are insensitive to the radial dependence and there is a priori no justification to the choice of radial function for $A$ and $\bA$ in \eqref{eq:aba}.

 
 

\section{Eternal black holes}\label{sec:defBH}

In general relativity, a Lorentzian eternal black hole can be maximally extended to possess two asymptotic regions that are connected through an Einstein-Rosen bridge. In the context of (ordinary, spin-$2$) AdS/CFT this is well-understood \cite{Israel:1976ur,Maldacena:2001kr}: the two asymptotic regions correspond to two copies of the dual field theory, and the black hole defines a thermofield state in the doubled field theory:
\be\label{eq:thf1}
|\psi \rangle = \frac{1}{\sqrt{Z}}\sum_n e^{-\frac{\beta}{2}\le(E_n + \mu Q_n\ri)} |{\cal U} n\rangle_L \otimes |n \rangle_R~.
\ee
We included in the definition of $|\psi \rangle$ a chemical potential $\mu$ that couples to a conserved charge $Q$ that commutes with $H$. Here $|n\rangle$ runs over a full basis of energy eigenstates of the CFT, $E_n$ and $Q_n$ labels their energies and charges, and $\sU$ is the anti-unitary operator that implements CPT. The full Hilbert space is composed by two copies of the original CFT Hilbert space: ${\cal H}= {\cal H}_L\otimes {\cal H}_R$.

We briefly review a Euclidean path integral ``explanation'' of this fact \cite{Maldacena:2001kr}. Consider performing the field theory Euclidean path integral on a manifold that is the product of the spatial direction(s) and an interval of Euclidean time with length $\frac{\beta}{2}$. It is necessary to specify field-theoretical boundary data on the two endpoints of the interval; the dependence of the path integral on the boundary data defines a state in the doubled copy of the field theory. This state is precisely \eqref{eq:thf1}. The suppression by $\exp\le(-\frac{\beta H}{2}\ri)$ arises from the evolution through $\frac{\beta}{2}$ of Euclidean time. 

Now consider implementing this procedure holographically. The path-integral over a full cycle of Euclidean time $\beta$ with periodic boundary conditions corresponds to studying the usual Euclidean black hole described above. We may however cut open this path integral after evolution through Euclidean time $\frac{\beta}{2}$ and analytically continue to Lorentzian time. The resulting Lorentzian manifold is the eternal maximally extended black hole, and the arguments above indicate that the resulting field-theory state is the thermofield state \eqref{eq:thf1}.  

Thus we expect that regular Euclidean gauge connections should (upon analytic continuation) map in a straightforward manner to the dual field theory in a thermofield state. This has consequences: as we review in Appendix \ref{app:kms}, 2-point functions on this state satisfy very specific periodicity conditions. Consider a charged scalar operator $\sO$, and we denote $\sO_L$ as an operator acting on  ${\cal H}_L$ and similarly for $\sO_R$. Two point functions that involve  $\sO_{L,R}$ satisfy
\bea\label{eq:kmscorr}
\langle \psi| \sO_{R}(t_f) \sO_{R}(t_i)|\psi \rangle &=& \langle \psi| \sO_{R}(t_f) \sO_{R}(t_i-i\beta)|\psi \rangle \cr &=&
\langle \psi| \sO_{L}(-t_f) \sO_{L}(-t_i) |\psi \rangle\cr &=&
\langle \psi| \sO_{L}(-t_f-i\beta/2) \sO_{R}(t_i) |\psi \rangle \cr &=& \langle\psi| \sO_{R}(t_f) \sO_{L}(-t_i-i\beta/2) |\psi \rangle\,.
\eea
In a mild notational abuse, we will refer to these all as Kubo-Martin-Schwinger or KMS conditions (even though technically only the first is ``the'' KMS condition). 

We may now ask whether relations such as \eqref{eq:kmscorr} are satisfied for eternal black holes in higher spin gravity. One immediate technical obstruction is that it is difficult to couple matter to these theories: a procedure as simple as probing the bulk with, for example, a scalar operator is cumbersome. This was one reason why in \cite{Kraus:2012uf} the question of the thermofield state was phrased in Vasiliev's higher spin gravity which includes a massive scalar field. 

However, this is an obstruction that we can now overcome. The recent developments in \cite{Ammon:2013hba,deBoer:2013vca,deBoer:2014sna,Hegde:2015dqh} show that a Wilson line operator is precisely the probe we need: it is a bulk observable that computes correlation functions of light operators in the dual CFT.  More concretely, we will consider
\be\label{eq:wilson1}
W_\sR(y_i,y_j)= \langle U_i | {\cal P} \exp\le( \int_{C_{ij}} A\ri){\cal P} \exp\le(\int_{C_{ij}} \bar A\ri) | U_f \rangle ~,
\ee
where $C_{ij}$ is a curve with bulk endpoints $(y_i,y_j)$ and $\sR$ is an infinite dimensional representation of the gauge group.  $U(y)$ is a probe field which lives on the wordline $C_{ij}$: its quantum numbers are governed by $\sR$ and it satisfies suitable boundary conditions at the endpoints (which we discuss in appendix \ref{app:wilson}). The key property is that as we take the endpoints to the boundary, the Wilson line gives \cite{deBoer:2014sna}
\be\label{eq:wcr}
W_\sR(y_i,y_j) \underset{r\to \infty}{=} \langle \Psi| \sO (x_i) \sO(x_j)|\Psi \rangle~.
\ee
Here  $(x_i,x_j)$ are boundary positions. $\sO (x_i)$ is an operator with scaling dimension $\Delta_{\sO}$ that is fixed as the central charge $c$ goes to infinity:\footnote{Or equivalently, in gravity we would say that it is a particle with a small mass in Planck units. In a rather crude way, we can identify $\sO$ with the probe field $U$. In this language, the Casimir's of the representation $\sR$ control the quantum numbers of the dual operator.} this is what we define as a `light' operator. The state $|\Psi\rangle$ is  `heavy', $\Delta/c$ is fixed as $c\to \infty$, and it corresponds to the background state created in the bulk by $(A,\bar A)$. 

\begin{figure}
\begin{center}
\includegraphics[width=0.5\textwidth,page=2]{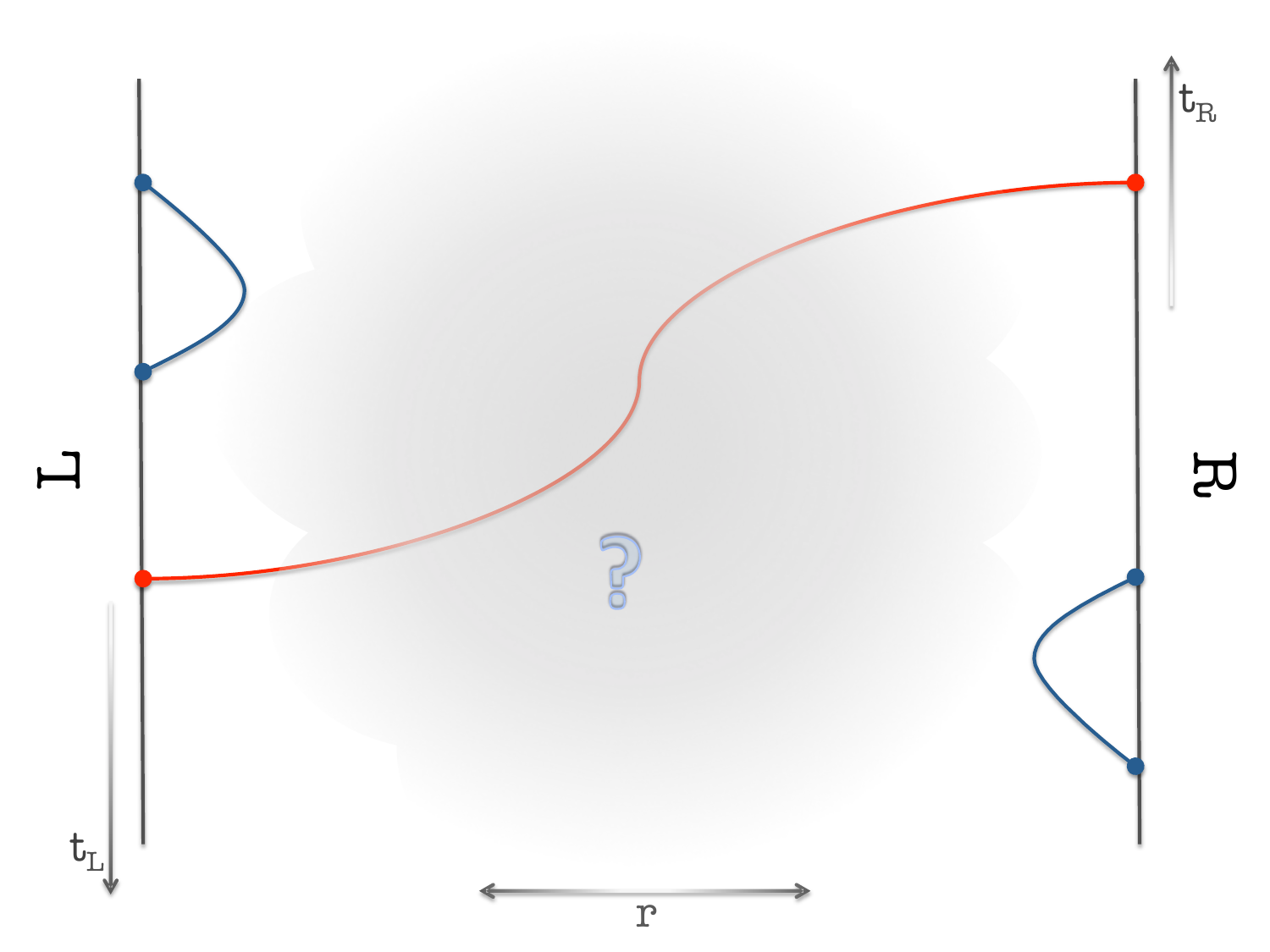}
\end{center}
\caption{Topology of the eternal black hole, which  contains at least two boundaries: the right ($R$) boundary at $r\to \infty$ and left ($L$) boundary at $r\to-\infty$. At this stage the interior is undetermined (and hence the question mark). The different lines correspond to various Wilsons lines we will study: blue lines correspond to $W_\sR(t_{i|R}, t_{f|R})$ or $W_\sR(t_{i|L}, t_{f|L})$, and red to $W_\sR(t_{i|R}, t_{f|L})$.}\label{fig:lor1}
\end{figure}

We will often be interested in the particular case when $|\Psi\rangle$ is the thermofield state \eqref{eq:thf1}: in that case we access operators in the left or the right tensor factor of the Hilbert space by taking the bulk points $y_i$ to the appropriate boundary. We will  omit explicit mention of a radial coordinate and use a subscript notation to indicate on which side the corresponding boundary coordinate is located. For example, for a correlator between the right and left boundary we have
\be
W_{\sR}(x_{i|R},x_{f|L}) = \langle \psi| \sO_R(x_i) \sO_L(x_f)|\psi\rangle~,
\ee
with $|\psi\rangle$ the thermofield state. A schematic depiction of the configutations we will study are shown in Figure \ref{fig:lor1}.

As we review in Appendix \ref{app:wilson}, the objects that controls the Wilson line are traces of the following matrix 
\bea\label{eq:M1}
M(y_i,y_f)=R(y_i)L(y_i)L^{-1}(y_f)R^{-1}(y_f)~,
\eea
which assumes that the connections are flat, i.e.  
\be
A=LdL^{-1}~,\quad\bar A= R^{-1}dR~. \label{nothingness}
\ee
While the Wilson line does in general transform under gauge transformations \eqref{bulkgauge} with support at its endpoints, it is invariant under the Lorentz subgroup of such transformations. 

The Wilson line gives us a fairly sensitive probe of higher spin geometry, allowing us to directly evaluate correlation functions such as those appearing in \eqref{eq:kmscorr}. As we will see, establishing the validity of relations such as \eqref{eq:kmscorr} in a two-sided black hole in the Chern-Simons formulation of gravity will require a more careful definition of Euclidean regularity than the holonomy condition \eqref{smoothness}.

\subsection{Refined notions of Euclidean regularity}\label{sec:refeuc}
Here we describe the conditions required for a thermofield interpretation. Consider first choosing a radial coordinate $r$ so that we can form the Cartesian complex coordinates
\be
w = r \exp\le(\frac{2\pi i \tau}{\beta}\ri)~, \qquad \bw = r \exp\le(-\frac{2\pi i \tau}{\beta}\ri) \ . 
\ee
We now claim that entirely regular physics on the Lorentzian section of a Euclidean black hole background -- i.e. the interpretation of in terms of a thermofield state -- requires {\it that the spacetime-dependent gauge parameters $L(y)$, $R(y)$ be smooth functions of $w, \bw$ near the Euclidean origin}. In particular, we will allow only non-negative integer powers of $w, \bw$ in a Taylor expansion about the origin:
\be
L,R(w,\bw \to 0) \sim \sum_{m,n \in \mathbb{Z^{+}}} c_{mn} w^{m} \bw^{n}~. \label{analyticity}
\ee

This is just the usual condition for smoothness of a scalar function at the origin of a disc $D^2$: nevertheless, interpreted from the Chern-Simons point of view, it is a stronger constraint on the bulk gauge connections than those normally considered in the literature. In particular, it is stronger than the holonomy condition \eqref{smoothness} in that it involves radial dependence as well as the Euclidean time direction. This same important observation was made recently in \cite{Banados:2016nkb}. The difference in the following will be the implementation of this more refined notion of regularity: the authors in \cite{Banados:2016nkb} considered directly the metric-like fields and our implementation uses solely the Chern-Simons connections. 

We will say that a connection satisfying \eqref{analyticity} is in {\it strong Kruksal gauge}: as we explain, it is the gauge-theoretical analog of the Kruskal coordinate system that permits passage through the horizon. Note that in this gauge we have
\be
A_{\tau}(r = 0) = L\p_{\tau}L^{-1}\big|_{r = 0} = r\le( \frac{2 \pi i}{\beta} L\le(e^{\frac{2\pi i\tau}{\beta}}\p_{w} -  e^{-\frac{2\pi i \tau}{\beta}}\p_{\bw}\ri)L^{-1}\ri)\bigg|_{r = 0} = 0~,
\ee
where the the smoothness condition \eqref{analyticity} ensures that the derivatives are regular at the origin, establishing the last equality. The time components of all gauge fields are zero at the horizon.  This is a very natural condition for gauge fields propagating on black hole background (and indeed is extensively used in the usual understanding of the thermodynamics of charged black holes --see e.g. \cite{Chamblin:1999tk,Hartnoll:2009sz}). It is thus interesting to note that the BTZ black hole written in the usual choice of gauge --despite \eqref{eq:aba} being widely accepted as being ``regular''-- actually does not satisfy it. 

There is, however a weaker gauge condition that one can impose. We see from \eqref{eq:M1} that the Wilson lines studied in this paper depend only on  the combination $R(y)L(y)$. Thus if we only care about such Wilson lines we might demand only that the composite field $R(y) L(y)$ be smooth as a function of $w,\bw$, and not the individual functions $R(y)$ and $L(y)$ themselves. We will call this {\it weak Kruskal gauge}. In weak Kruskal gauge we find only that $A_{\tau} - \bA_{\tau} = 0$ at the horizon, and the usual BTZ black hole turns out to already be in weak Kruskal gauge. We note that while the Wilson lines discussed in this paper cannot tell the difference between strong and weak Kruskal gauges, other probes that couple less symmetrically to the left and right connections -- such as e.g. a particle with spin\footnote{See \cite{Castro:2015csg,Castro:2014tta} for work towards constructing a Wilson line to describe such a particle.} -- will be sensitive to the difference, and we expect such probes to display regular behavior only in strong Kruskal gauge. Importantly, the higher spin black hole as written in \eqref{bhconn} is {\it not} in either Kruskal gauge. 

The need for such conditions is most easily understood with a toy model of a flat $U(1)$ gauge field $B$ in two dimensions. As a proxy for the near-horizon region, consider Euclidean $\mathbb{R}^2$:
\be
ds^2 = dr^2 + r^2 d\tau^2 = dw d\bw~,
\ee
with $w = r e^{i\tau}$ as usual. As $B$ is flat, it can be written in terms of a group element $g(w,\bw) \in U(1)$:
\be
B = g^{-1} dg~,
\ee
This is the $U(1)$ analog of \eqref{nothingness}. The $U(1)$ analog of the holonomy condition \eqref{smoothness} merely states that $g$ should be single-valued around the $\tau$ circle, i.e 
\be
g(r, \tau + 2\pi) = g(r,\tau)~. \label{u1smoothness}
\ee
Importantly, it makes no reference to the radial direction. In particular, consider e.g. 
\be
g_0(r,\tau) = e^{i\tau} = \sqrt{\frac{w}{\bw}}\ , \label{badguy}
\ee
which respects this holonomy condition. Consider now a particle with $U(1)$ charge $q$ moving on this Euclidean background: its action contains a term $iq\int_C B$ integrated along its worldline $C$, and there are no obvious pathologies associated with it.

 \begin{figure}
\begin{center}
\includegraphics[width=0.5\textwidth,page=4]{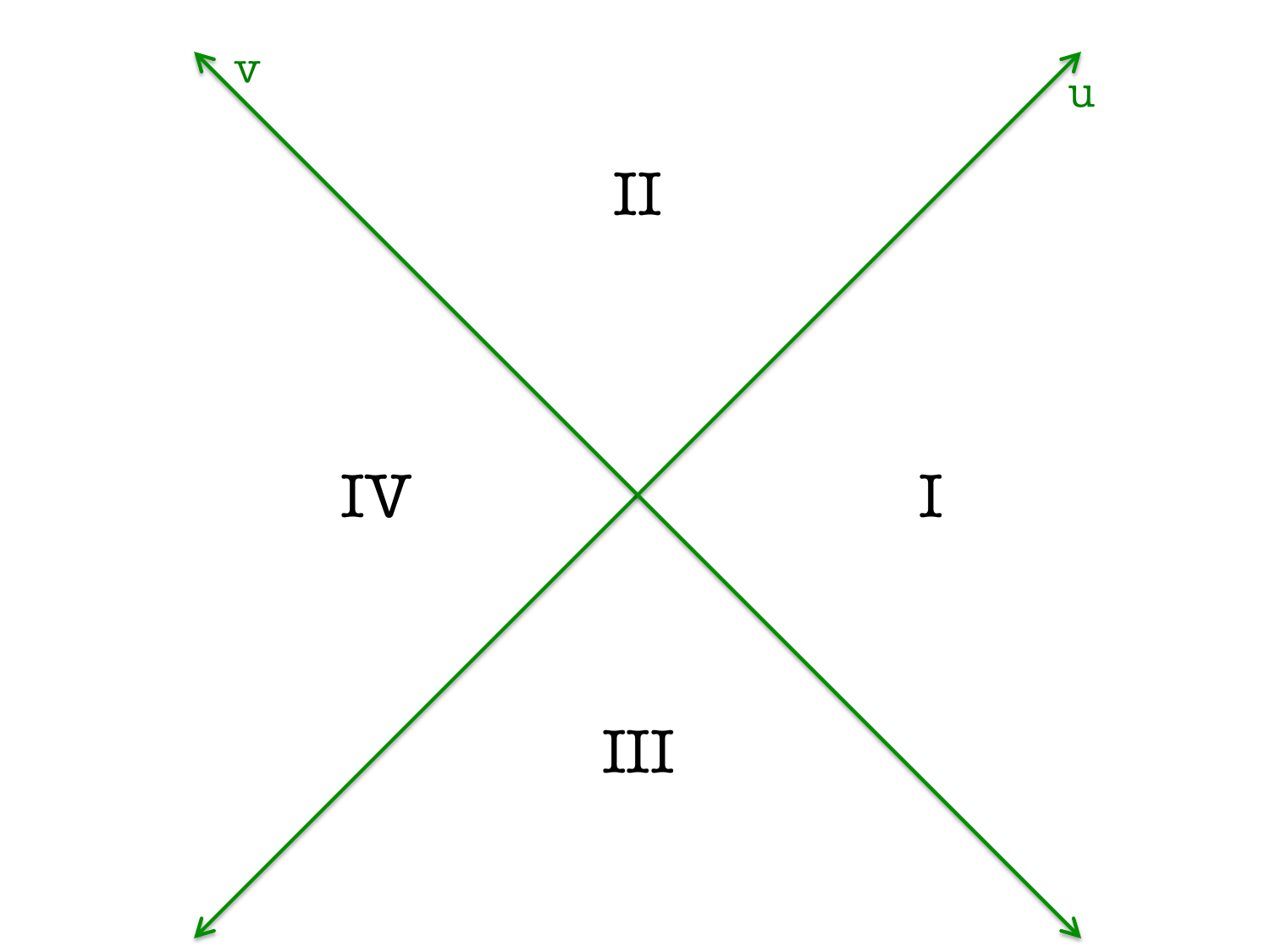}
\end{center}
\caption{Four quadrants covered by the coordinates $(u,v)$ on $\mathbb{R}^{1,1}$. In quadrant {\bf I}: $u>0$, $v<0$;  in quadrant {\bf IV}: $v>0$, $u<0$. }\label{fig:uv}
\end{figure}

Now we analytically continue Euclidean $\mathbb{R}^2$ to Rindler space $\mathbb{R}^{1,1}$ in the usual way via
\be
w = -v \qquad \bw = u. \label{anncont}
\ee The Lorentzian metric is simply $ds^2 = -du dv$ and is well defined for all $u,v$ in all quadrants in Figure \ref{fig:uv}. However we now find that the analytic continuation of the $U(1)$ group element is
\be
g_0(u,v) = \sqrt{-\frac{u}{v}} \ . 
\ee  
This has a branch cut along the horizons $uv = 0$: in other words, without specifying more information, the phase acquired by a charged particle moving on the Lorentzian section is ill-defined as we cross the quadrants in Figure \eqref{fig:uv}. Thus the innocuous-seeming Euclidean group element \eqref{badguy} does not result in well-defined Lorentzian physics. The $\sln$ analog of this pathology will manifest itself later on when we attempt to compute two-sided correlators in the eternal black hole and demonstrate consistency with the properties of the thermofield state. 

Precisely to avoid such ambiguities when performing the analytic continuation \eqref{anncont}, the Kruskal gauge condition demands that $g(x)$ -- or rather its $\sln$ analogs $L(y)$ and $R(y)$ -- be smooth functions of $w$, $\bw$, and thus also of $u, v$ after analytic continuation. 

We now show that the relations \eqref{eq:kmscorr} follow from weak Kruskal gauge. As described in \eqref{eq:M1}, boundary theory correlation functions are controlled through the Wilson line by the object
\bea\label{eq:M1b}
M(y_i,y_f)=R(y_i)L(y_i)L^{-1}(y_f)R^{-1}(y_f)~,
\eea
where the points are at one of the two boundaries. We need to understand how to go from Euclidean to Lorentzian time: as is conventional, the mapping is
\be
w = f(r) e^{\frac{2\pi i}{\beta}\tau} \rightarrow -v~, \qquad \bw = f(r) e^{-\frac{2\pi i}{\beta}\tau} \rightarrow  u~, \label{wuv}
\ee
where $f(r)$ is an {\it odd} function that vanishes linearly at the black hole horizon and diverges at the AdS boundary. In  quadrant {\bf I} we have $u > 0$ and $v < 0$, which we parametrize in terms of a Lorentzian time coordinate $t_R$ as
\be
u = f(r) e^{\frac{2\pi}{\beta}t_R}~, \qquad v = - f(r)e^{-\frac{2\pi}{\beta} t_R} ~.\label{R1}
\ee
In quadrant {\bf IV} we have $u < 0$ and $v >0$, which we parametrize as
\be
u = -f(r) e^{-\frac{2\pi}{\beta}t_L}~, \qquad v =  f(r)e^{\frac{2\pi}{\beta} t_L} \label{L1} \ . 
\ee
This identification uniquely fixes $M$ in the entire maximally extended spacetime. We may now verify the validity of the relations \eqref{eq:kmscorr}, which require that the two-point function $\langle\psi| \sO_{R}(t_f) \sO_{R}(t_i)|\psi\rangle$ is equal to all of the following:
\begin{enumerate}
\item $\langle \psi| \sO_{R}(t_f) \sO_{R}(t_i-i\beta)|\psi \rangle$.
A shift in $\tau$ by a full period $\beta$ has no action on $u,v$, $u \to e^{2\pi i}u$, and $M$ is single-valued as a function of $u,v$. This property (and only this property) actually follows from the holonomy condition \eqref{smoothness} alone and does not require a Kruskal gauge. 
\item $\langle \psi| \sO_{L}(-t_f-i\beta/2) \sO_{R}(t_i) |\psi \rangle$. From the global coordinates \eqref{L1} and \eqref{R1} we see that the point labeled by $(r = r_{\Lam},t_R = t_f)$ in the right quadrant is the {\it same} as the point labeled by $\le(r = r_{\Lam}, t_L = -t_f \pm i \frac{\beta}{2}\ri)$ in the left quadrant. Taking $r_{\Lam} \to \infty$ now relates $M$ to the appropriate correlation function. The equality with $\langle \psi| \sO_{L}(-t_f-i\beta/2) \sO_{R}(t_i) |\psi \rangle$ follows in the same way. 
\item $\langle \psi| \sO_{L}(-t_f) \sO_{L}(-t_i) |\psi \rangle$. This equality is most easily understood by moving each point from the right quadrant to the left using the manipulation above, and then translating both arguments in Euclidean time by $\frac{i\beta}{2}$. 
\end{enumerate}
These relations may seem like kinematic trivialities: however it is important to note that if we do not pick the bulk gauge connections to satisfy \eqref{analyticity}, then branch cuts in the $u,v$ plane mean that the relations above do not hold -- for example the second relation was not satisfied by the scalar field correlators computed in \cite{Kraus:2012uf}. We believe that \eqref{analyticity} are, however, crucial for a complete interpretation of the black hole as a thermofield state. 

\subsection{Parametrizing black hole connections in Kruskal gauge}
Having established the desirable properties of these gauges, we now turn to their explicit construction. As it turns out, any black hole can be placed in (either strong or weak) Kruskal gauge. Recall from \eqref{eq:aba} that the standard parametrization of black hole solutions to Chern-Simons gravity involves two constant flat connections $a, \ba$ that point only in the field theory directions, in terms of which \eqref{nothingness} becomes
\begin{align}\label{RL0}
 L(y)= b(r)^{-1}\,\exp\left(-\int^x_{0}dx^i {a}_i\right)\, ,\quad R(y)=\text{exp}\left(\int^x_{0}dx^i \bar{a}_i\right)\bar b(r)^{-1}\, , 
\end{align}
where we have generalized slightly by allowing for a different radial function for the barred and unbarred coordinates; $x^i$ runs only over field theory coordinates. 

From here we find that $A_{\tau}(r) = b^{-1}(r) a_{\tau} b(r)$ and thus is never zero for any value of $r$. This presentation of the black hole is then not in strong Kruskal gauge. To put it into strong Kruksal gauge, we will need to ``unwrap'' the effect of moving in $\tau$.  Note that \eqref{smoothness} tells us that $a_{\tau}$ and $\ba_{\tau}$ are conjugate to $L_0$
\be
a_{\tau} = V \le(\frac{2 \pi i L_0}{\beta} \ri)V^{-1}~, \qquad \ba_{\tau} = \bV \le(\frac{2 \pi i L_0}{\beta}\ri) \bV^{-1} \ . \label{diagA}
\ee
Consider now the following gauge transformation:
\be
L^{(K)} = \Lam_L L ~,\qquad R^{(K)} = R \Lam_R ~,\qquad \Lam_L = \Lam_R^{-1} = \exp\le(\frac{2\pi i L_0}{\beta}\tau\ri)G~. \label{kruskalify}
\ee
Here $G$ is a constant (arbitrary) element of the group. The gauge transformed connections can be written
\be
A^{(K)} = B(r,\tau)^{-1}\le(a_{(\phi)} + d\ri) B(r,\tau) ~,\qquad \bA^{(K)} = \bB(r,\tau)\le(\ba_{(\phi)}+d\ri)\bB^{-1}(r,\tau)~. \label{Bform}
\ee
Here the notation indicates that $a_{(\phi)}$ is a connection whose $\phi$ component is equal to that of the original $a$ but whose $\tau$ component is zero. We have
\be
B(r,\tau) = e^{a_{\tau} \tau} b(r)G^{-1} e^{-i L_0 \frac{2\pi \tau}{\beta}}~, \qquad \bB(r, \tau) = e^{i L_0 \frac{2 \pi \tau}{\beta}}G\bb(r)e^{- \ba_{\tau}\tau}~. \label{Bexp}
\ee
The gauge transformation \eqref{kruskalify} is far from unique. There are two crucial features of our choice. First, it is important that it winds once around $\slt$ as we traverse the time cycle. Second, it is a Lorentz transformation: this assures that the gauge transformation does not affect \eqref{eq:M1} which evaluates CFT correlators.

In the gauge \eqref{Bform} we can now impose the strong Kruskal gauge condition \eqref{analyticity}. Focusing for now on the unbarred connection, we see that this new parametrization treats $r$ and $\tau$ together in the new object $B(r,\tau)$. It is convenient to use \eqref{diagA} to rewrite
\be
B(r,\tau) = V e^{\frac{2\pi i \tau}{\beta}L_0} V^{-1} b(r) G^{-1} e^{-i\frac{2\pi \tau}{\beta}L_0}~.
\ee
The full $\tau$ dependence now enters in the conjugation of $V^{-1} b(r)G^{-1}$ by $e^{i\frac{2\pi \tau}{\beta}L_0}$. The smoothness condition \eqref{analyticity} tells us that in the expansion of $B$ around the origin we can only have terms of the form $r^n e^{\pm\frac{ 2\pi i n \tau}{\beta}}$ with $n$ integer, thus tying together the $r$ and $\tau$ dependence. This is a constraint on $b(r)$: given a choice of $a$, we can now explicitly solve for $b(r)$. Typically we demand that $b(r)$ approach the standard choice at infinity so that our connections satisfy asymptotically AdS boundary conditions.\footnote{It is very important that $b(r)$ and $\bb(r)$ asymptote $e^{r L_0}$ as $r \to \infty$. Relations such as \eqref{eq:wcr} rely on this profile at infinity, and we do not want to tamper with it.}  We note that there is still considerable freedom in the choice of $b(r)$: its behavior at infinity and at the horizon is fixed, but the topological nature of the theory means that it is essentially utterly unconstrained in the interior. In Appendix \ref{app:kruskal} we demonstrate an algorithm to find a suitable $b(r)$ explicitly for the higher spin black hole. 

We turn now to weak Kruskal gauge. Here there is no need for an ``unwrapping'' procedure: instead, we may start from the original \eqref{RL0} and using the explicit diagonalization \eqref{diagA} we find
\be
R(y) L(y) = \bV\exp\le(\frac{2 \pi i L_0\tau}{\beta}\ri)\bV^{-1}\le(\bb(r)^{-1} b(r)^{-1}\ri) V\exp\le(-\frac{2 \pi i L_0\tau}{\beta}\ri)V^{-1}\, ,
\ee
where we have omitted the $\phi$ dependence. We see that it is now the object $\bV^{-1}\le(\bb(r)^{-1} b(r)^{-1}\ri) V$ that is conjugated by $e^{i\frac{2\pi \tau}{\beta}L_0}$: thus the analyticity condition applied to $R(y)L(y)$ can be viewed as a weaker condition on the product $b(r)\bb(r)$. 

To summarize: to put a black hole into weak Kruskal gauge we only need to judiciously choose the {\it product} $b\bb$. To put it into strong Kruskal gauge we must unwrap the $\tau$ dependence via a Lorentz transformation and then judiciously choose $b(r), \bb(r)$. 


\section{Eternal BTZ in Chern-Simons formulation} \label{sec:btz}

In this section we warm up by studying the familiar BTZ black hole in the Chern-Simons formulation of $\slt$ gravity. We will demonstrate that the definitions above permit access to all regions of the maximally extended spacetime. The results here can be compared with those obtained from the usual metric description of the BTZ black hole; see e.g. \cite{Banados:1992gq,Maldacena:2001kr,Kraus:2002iv} 

The metric of the non-rotating BTZ black hole can be written
\be
ds^2 = -e^{-2\rho}\le(e^{2\rho} - {2 \pi \sL}\ri)^2 dt^2 + e^{-2\rho}\le(e^{2\rho} + {2\pi\sL}\ri)^2 d\phi^2 + d\rho^2~. \label{origBTZ}
\ee 
The corresponding connections can be written in the notation introduced in \eqref{eq:aba}:
\begin{equation}
A= b(r)^{-1}\le(a(x^+,x^-) + d\,\ri)b(r)\,,\qquad \bar{A} = \bb(r)\le(\bar{a}(x^+,x^-) + d\,\ri)\bb(r)^{-1}\,.
\end{equation}
where we have
\bea\label{bhconnBTZ}
 a = \le(L_1-{2 \pi  \mathcal{L}}L_{-1}\ri) dx^+\, ,\qquad \bar{a} =-\left(L_{-1}-{2 \pi  \mathcal{L}}L_1\ri) dx^- \ .
 \eea
The black hole temperature can be determined by imposing the holonomy condition \eqref{smoothness} and is $\beta =  \sqrt{\frac{\pi}{2  \sL}}$. In the literature there is a standard choice for the radial functions $b(r)$, $\bb(r)$: in this section we will instead {\it derive} them by demanding Euclidean regularity in the sense described in the previous section. 
The gauge connections \eqref{bhconnBTZ} can be diagonalized as in \eqref{diagA}. The definition of the similarity matrices $V, \bV$ leaves unfixed the normalizations of each of the eigenvectors. By adjusting these normalizations $V, \bV$ can be made to have unit determinant and also satisfy the following relations:
\be
V(L_1 - L_{-1})V^{-1}  =  -2 L_0 ~,\qquad \bV(L_1 - L_{-1})\bV^{-1}  =  - 2 L_0~, \label{conjop}
\ee
as well as be related to each other via
\be
V\bV^{-1} =  \exp(2 \rho_0 L_0)~, \quad \rho_0 \equiv \ha \log\le({2 \pi \sL}\ri) \ . \label{magic}
\ee
The relations among $V$ and $\bar V$ -- which are unique to $sl(2)$ and do not have a simple analog in the higher spin case -- permit simple computations to be performed in the BTZ case. $\rho_0$ has been presciently named, but at this moment has no geometric significance. 

\subsection{Strong Kruskal gauge}

We would first like to put the connections \eqref{bhconnBTZ} in strong Kruskal gauge. We perform a time-dependent Lorentz transformation of the form described in \eqref{kruskalify}:
\be
\Lam_L = \Lam_R^{-1} = \exp\le(\frac{2\pi i L_0}{\beta} \tau\ri) V^{-1} e^{\rho_0 L_0}\ . 
\ee
With the benefit of hindsight, we have chosen  $G=V^{-1} e^{\rho_0 L_0}$. Using \eqref{magic} this is equivalent to
\be
\Lam_L = \Lam_R^{-1} = \exp\le(\frac{2\pi i L_0}{\beta} \tau\ri) \bV^{-1} e^{-\rho_0 L_0} \ . \label{barredlor}
\ee
We now find that the gauge-transformed connection in the unbarred sector takes the form \eqref{Bform} with
\be
B^{-1}(r,\tau) = e^{\frac{2\pi i L_0}{\beta} \tau}V^{-1} e^{\rho_0 L_0} b^{-1}(r) V e^{-\frac{2\pi i L_0}{\beta} \tau} V^{-1},
\ee
Consider now the Euclidean coordinates:
\be
w \equiv \tanh\le(\frac{r}{2}\ri) e^{\frac{2\pi i}{\beta} \tau} ~,\qquad \bw = \tanh\le(\frac{r}{2}\ri) e^{-\frac{2\pi i}{\beta} \tau} ~.\label{wdef}
\ee
Here (again with the benefit of hindsight) we have picked a specific radial function $\tanh\le(\frac{r}{2}\ri)$ of $r$: in order for this change of coordinates to be well-defined this function must be odd and have a smooth Taylor expansion in odd powers of $r$ (starting with the linear term in $r$) near $r = 0$. We now demand that $B(r,\tau)$ be a smooth function of $w,\bw$. This is conveniently viewed as a constraint on the function $V^{-1} e^{\rho_0 L_0} b^{-1}(r) V$. 

We briefly digress from this specific example to discuss the general case: consider expanding 
\be
V^{-1} e^{\rho_0 L_0} b^{-1}(r) V = \exp\le(\sum_a F_a(r) T^a\ri)~,
\ee
with the $F_a(r)$ a set of mode functions and the $T^a$ running over the generators of the algebra. The conjugation by $e^{\frac{2\pi i L_0}{\beta} \tau}$ attaches a power of $e^{-\frac{2\pi i h_a}{\beta} \tau}$ to each term in the sum, where $h(a)$ is the weight of the generator $T^a$ under $L_0$. The analyticity condition then requires that $F_a(r \to 0) \sim r^{|h(a)|}$, so that the full radial and time dependence can be expressed as a product of integer powers of $w$ and $\bw$. In the higher spin case this system of constraints must be systematically solved, as explained in Appendix \ref{app:kruskal}. 

However for the purposes of the BTZ black hole it is sufficient to make a rather simple and consistent choice for $F_a$; we can take 
\be
V^{-1} e^{\rho_0 L_0} b^{-1}(r) V = \exp\le(\frac{r}{2}\le(L_1 - L_{-1}\ri)\ri)~.
\ee
This choice satisfies the condition above, as $L_{\pm 1}$ have weight $\pm 1$. Using \eqref{conjop} we then find
\be
b(r) = \exp\le((r + \rho_0) L_0\ri)\ . 
\ee
We can follow precisely the same procedure for the barred sector (using now the form of the gauge transformation in \eqref{barredlor}) to derive an expression for $\bb(r)$ and conclude that $\bb(r) = b(r)$. 

Finally, to put this into a more familiar form we can define a new coordinate $\rho \equiv r + \rho_0$, in terms of which we have
\be
b(\rho) = \bb(\rho) = e^{\rho L_0}
\ee
This is of course the usual choice of radial gauge function for the $\slt$ gravity, which we have now derived. Note that the horizon -- which has physical significance as the fixed point of translations in Euclidean time, and the place where the time components of the Kruskal connections vanish -- is at $r = 0$, which maps to the usual $\rho = \rho_0$. In this approach $\rho_0$ appeared purely algebraically from the original relation \eqref{magic}. 

\subsection{Maximally extended  connections}
From above we can now explicitly compute the spacetime-dependent gauge parameters $L(y)$ and $R(y)$ on the Euclidean section in the strong Kruskal gauge that we have constructed: in terms of $w,\bw$ in \eqref{wdef} we find
\begin{align}
L(y) & = \frac{1}{\sqrt{2\pi\beta(1-w \bw)}} \left(
\begin{array}{cc}
 e^{-\frac{\pi  \phi }{\beta }} \left(e^{\frac{2 \pi  \phi }{\beta }}-w\right) \beta  & e^{-\frac{\pi  \phi }{\beta }}  \left(w+e^{\frac{2 \pi  \phi }{\beta }}\right)\pi  \\
 e^{-\frac{\pi  \phi }{\beta }} \left(\bw e^{\frac{2 \pi  \phi }{\beta }}-1\right) \beta  & {e^{-\frac{\pi  \phi }{\beta }} \left(e^{\frac{2 \pi  \phi }{\beta }} \bw+1\right)\pi  } \\
\end{array}
\right) ~,\cr
R(y) & = \frac{1}{\sqrt{2\pi\beta(1 - w \bw)}}
\left(
\begin{array}{cc}
 e^{-\frac{\pi  \phi }{\beta }} \left(e^{\frac{2 \pi  \phi }{\beta }}-\bw\right) \beta  & e^{-\frac{\pi  \phi }{\beta }} \left(e^{\frac{2 \pi  \phi }{\beta }} w-1\right)\beta  \\
 {e^{-\frac{\pi  \phi }{\beta }} \left(\bw+e^{\frac{2 \pi  \phi }{\beta }}\right) \pi } & {e^{-\frac{\pi  \phi }{\beta }}   \left(e^{\frac{2 \pi  \phi }{\beta }} w+1\right)\pi} \\
\end{array}
\right)~.
\end{align}
They are analytic and smooth functions of $w, \bw$ near the origin. There is a singularity at $w\bw = 1$: from \eqref{wdef} we see that this is the AdS boundary. 

\begin{figure}
\begin{center}
\includegraphics[width=0.5\textwidth,page=3]{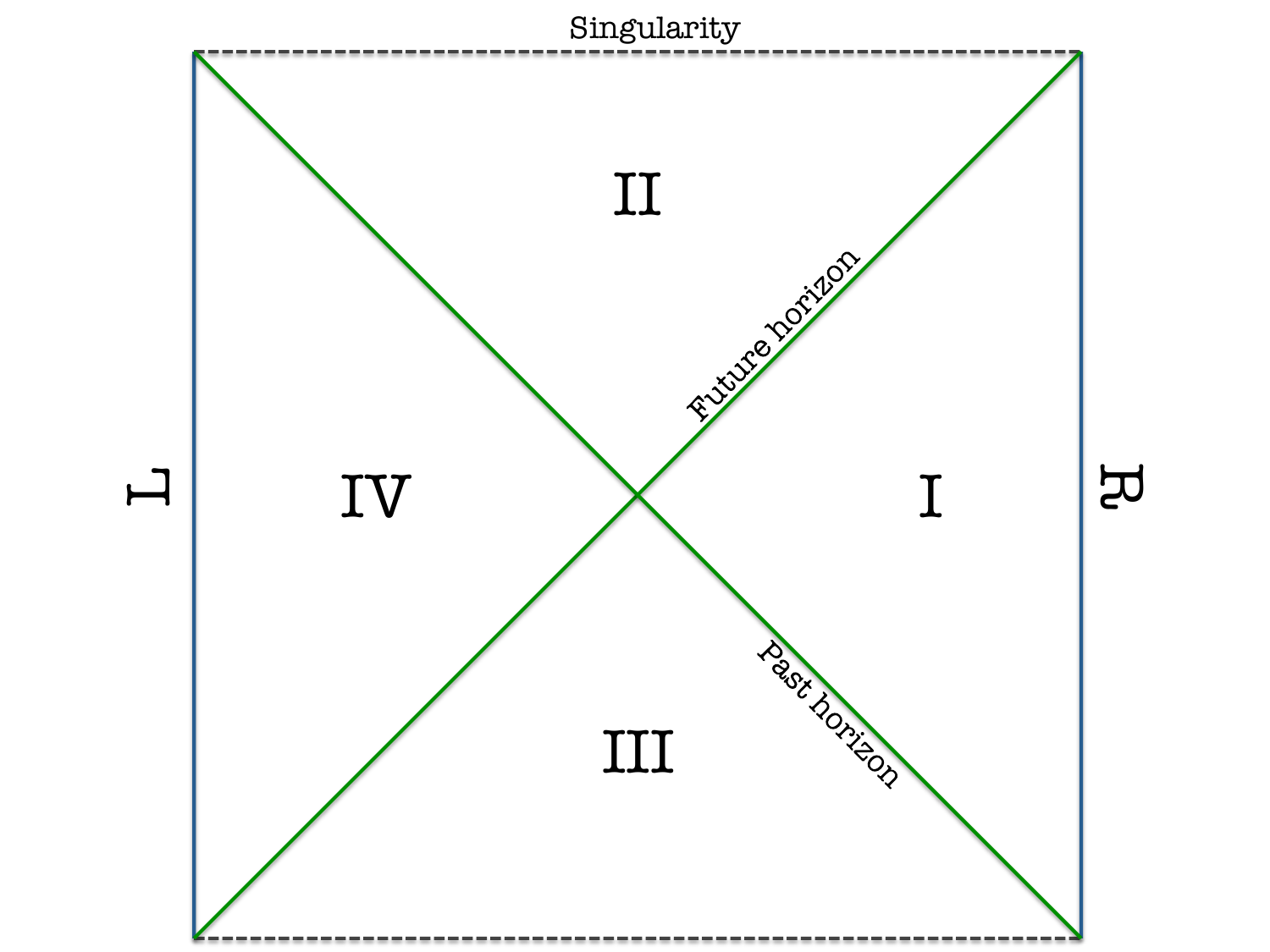}
\end{center}
\caption{Penrose diagram for static BTZ solution.}\label{fig:btz}
\end{figure}

We can now analytically continue to the real-time coordinates $u$ and $v$ via \eqref{wuv} to obtain gauge parameters that are well defined on the entire maximally extended spacetime. Though we do not need it, we may also compute the metric following from these connections: 
\be
ds^2 = -\frac{4}{(1+uv)^2} du dv + \le(\frac{2\pi}{\beta}\ri)^2 \le(\frac{uv - 1}{uv + 1}\ri)^2 d\phi^2~.
\ee
This is the usual BTZ metric in Kruskal coordinates, and the associated Penrose diagram is depicted in Figure \ref{fig:btz}. It is important to note that this is nothing but the coordinate transformation of the original BTZ metric \eqref{origBTZ}: the gauge transformation that we performed on the gauge connections to put it into strong Kruskal gauge is in the Lorentz subgroup of $\slt \times \slt$, and so does not affect the metric.  

From the form of $L(y)$ and $R(y)$ written above it is now straightforward to compute the Wilson line. Using \eqref{eq:M1} we compute between the trace of $M$ between any two points $(u_i, v_i)$ and $(u_f,v_f)$ (we set the spatial separation to $0$); this gives 
\be
\Tr(M(y_i, y_f)) = \frac{2}{(1+u_f v_f)(1+u_iv_i)}\le((1-u_i v_i)(1-u_fv_f) + 2 (u_f v_i + u_i v_f)\ri)~.
\ee 
 The Wilson line between two points is related to this object via
\be
\log W_{\sR} = -2h \cosh^{-1}\le(\ha \Tr(M)\ri)~.
\ee
Now by taking these points to the appropriate boundaries we may compute boundary correlators. It is instructive to map back to boundary time using the appropriate version of \eqref{R1} and \eqref{L1}: on quadrant {\bf I} with $u > 0$ and $v <0$ we have
\be
u = \tanh\le(\frac{r}{2}\ri)e^{\frac{2\pi t_R}{\beta}}~, \qquad v = -\tanh\le(\frac{r}{2}\ri)e^{-\frac{2\pi t_R}{\beta}}~,
\ee
and on quadrant {\bf IV} we have $u < 0$ and $v >0$, leading to
\be
u = -\tanh\le(\frac{r}{2}\ri)e^{-\frac{2\pi t_L}{\beta}} ~,\qquad v = \tanh\le(\frac{r}{2}\ri)e^{\frac{2\pi t_L}{\beta}} ~.
\ee
Note that the globally defined Killing vector corresponding to time translations is $u\p_{u} - v\p_{v}$, which is $\p_{t_R}$ on the right side and $-\p_{t_L}$ on the left side. 

Computing now the correlator between two points at the $R$ boundary and keeping track only of the universal information, we find
\be
\log W_{\sR}(t_{i|R},t_{f|R})= -2h \log\le(-\frac{1}{\ep^2}\sinh^2\le(\frac{\pi}{\beta}(t_f-t_i)\ri)\ri)~,
\ee
where we have introduced a UV cutoff $\ep$ that vanishes at $r \to \infty$. On the other hand, we may also compute a correlator between an initial point on the right boundary and a final point on the left boundary. We then find
\be
\log W_{\sR}(t_{i|R},t_{f|L}) = -2h \log\le(\frac{1}{\ep^2}\cosh^2\le(\frac{\pi}{\beta}(t_f + t_i)\ri)\ri)~.
\ee
These are of course just the usual results for geodesic distance on the BTZ black hole background. Note in particular that the two-sided correlator is related to the one-sided correlator by the KMS relation \eqref{eq:kmscorr}.

\section{Eternal higher spin black holes} \label{sec:hsbh}

In this section we study the Lorentzian properties of higher spin black holes solutions. In particular, we will consider three different gauges which differ only by the radial parametrization of the connection:
\begin{description}
\item[Wormhole gauge:] This corresponds to the choice of radial parametrization as $b(r)=\bar b(r)=e^{r L_0}$. The metric and connections are smooth for the entire range of $r$, with no horizon: hence it is a `wormhole'.  This gauge does not satisfy neither the weak or strong Kruskal condition. However, it does asymptote to AdS in the conventional sense at the  $R$ boundary, i.e. $r\to\infty$ in Figure \ref{fig:lor1}, and hence reproduces CFT correlators.
\item[Horizon gauge:] This gauge is designed to give a horizon in the metric of the higher spin black hole. An explicit constructions is given in  \cite{Ammon:2011nk}: this solution does satisfy the weak Kruskal condition, however it does not asymptote to AdS on either side of  Figure \ref{fig:lor1}. 
\item[Strong Kruskal gauge:] An explicit construction of connections that satisfies \eqref{analyticity} and reproduces correctly the dual CFT correlators.  
\end{description}

In higher spin gravity we lack the hindsight of BTZ due to the alternative metric formulation in the spin-2 case.  Our way to probe and test our definitions will be to use  the Wilson line \eqref{eq:wilson1} on the three Lorentzian backgrounds listed above.  As mentioned around \eqref{eq:wcr}, $W_\sR(C_{ij})$ captures boundary (CFT) correlators which allows us to test the KMS relations \eqref{eq:kmscorr} for arbitrary probes. More importantly,  $W_\sR(C_{ij})$ is the object that describes the dynamics of massive (charged or not) particles in Chern-Simons theory: this gives a robust definition of causality and connectedness  of the geometry which we can easily implement and exploit.  

For concreteness, we will focus on higher spin black holes in $sl(3)\times sl(3)$ Chern-Simons theory. For this theory we have
\be\label{eq:wr3}
-\log W_{\mathcal{R}}(C_{ij})=\Tr(\log(M) P_0)~, \quad P_0= {h\over 2}L_0 + {w_3\over 2} W_0~,
\ee
where $P_0$ governs the two quantum numbers of the representation $\sR$: $h$ which is the mass (or conformal dimension)  and $w_3$ corresponding to the spin-3 charge of the probe. More details can be found in Appendix \ref{app:wilson}. For $h\neq0 $ and $w_3=0$, equation  \eqref{eq:wr3} is the most natural definition of  `geodesic' in higher spin gravity; in particular, we will use the sign of $\log W_{\mathcal{R}}(C_{ij})$ to signal if endpoints are either spacelike, timelike or null separated. This is the key to associating a Penrose diagram to a given solution, and justify why our definition of Kruskal gauge actually gives rise to the desired definition of eternal black hole. 

\subsection{Failures and successes of the wormhole gauge}\label{sec:WH}

The wormhole gauge corresponds to black hole connections of the form \eqref{eq:aba} with  boundary components given by \eqref{bhconn}-\eqref{can} and radial functions $b(r)=\bar b(r)=e^{r L_0}$. This is the most commonly used parametrization of the connections in the literature. 

To probe the geometry we will evaluate $W_\sR(y_i,y_f)$ for the configurations shown in Figure \ref{fig:lor1}.  To start, we consider a Wilson line with no time separation $\Delta t=0$, and with both endpoints in the asymptotic region $R$: for both holomorphic \eqref{hol} and canonical \eqref{can} solutions, the result is 
 \begin{align}\label{WilsonRRx}
-\log W_\sR(x_{i|R}, x_{f|R})&= h \log \left(\frac{\beta  \sinh \left(\frac{\pi  \Delta \phi}{\beta }\right)}{ \pi \epsilon}\right)^4+\frac{12 h\mu^2}{\beta^2}\left[\frac{32\pi^2}{9}\left(\frac{\sigma\pi\Delta \phi}{\beta}\right)\text{coth}\left(\frac{\pi\Delta \phi}{\beta}\right)-\frac{20\pi^2}{9}  \right.\\
& \left. -\frac{4\pi^2}{3}\text{cosech}^2\left(\frac{\pi\Delta \phi}{\beta}\right)\left\{ \left(\frac{\sigma\pi\Delta \phi}{\beta}\text{coth}\left(\frac{\pi\Delta \phi}{\beta}\right)- 1  \right)^2   +\left(\frac{\sigma\pi\Delta \phi}{\beta}\right)^2   \right\} \right] +O(\mu^ 4) \,,\nonumber
 \end{align}
where we used \eqref{Wilsonm1m2} with  $w_3=0$ and expanded around $\mu\rightarrow 0$. Recall that in this notation $x_{i|R}$ denotes that the endpoint is placed at $r\to \infty$ (while  $x_{i|L}$ used below will refer to $r\to -\infty$).  The symbol $\sigma$ has been introduced to differentiate between the two types of black holes  
\bea\label{eq:sigma}
&&\sigma= 2: \textrm{holomorphic black hole~,} \cr
&&\sigma=1:  \textrm{canonical black hole~.} 
\eea
These are the results originally reported in  \cite{deBoer:2013vca,Ammon:2013hba}. 

The Wilson line has different features depending on whether the  holomorphic and canonical solution is used. When expanded to first order in $\mu$, the Wilson line \eqref{WilsonRRx} for the canonical black hole matches a perturbative CFT result found in \cite{Datta:2014ska} when $h=c/12(n-1)$: this corresponds to the dimension of the twist field that evaluates entanglement entropy as $n\to1$. When interpreted as entanglement entropy, strong subadditivity inequalities imply that the Wilson line must be nondecreasing and concave down as a function of $\Delta\phi$ \cite{Callan:2012ip}. Direct examination of the function above shows that this is true for the canonical black hole \cite{Knuttel:Thesis2014,Castro:2014mza}, but is not true for the holomorphic black hole \cite{deBoer:2013vca,Ammon:2013hba}. 

Another key requirement for the entanglement entropy is that when evaluated for large intervals in a mixed state, it should saturate to a linearly growing result $S_{EE}(\Delta\phi) \sim s \Delta\phi$ where $s$ is the ordinary thermal entropy density associated to the mixed state. For most values of $C$ this is true for both kinds of black hole, but for the holomorphic black hole there is an eigenvalue crossing at $C_0 = 3(9+\sqrt{33})/8\simeq 5.53$, and for $C < C_0$ the asymptotic limit of the holomorphic black hole entanglement entropy is then {\it not} consistent with its own thermal entropy density. While we present computations in both kind of black hole for completeness, we will restrict attention to the better-behaved canonical black hole when discussing the physical implications of our results. 

The above result  is only probing physics at the $R$ boundary in Figure \ref{fig:lor1}, but we can easily explore the properties of the geometry by moving the endpoints of the Wilson line. To start we set $\Delta \phi=0$ and explore the dependence on the $(r,t)$ plane. The Wilson lines for various configurations in Figure \ref{fig:lor1} reads
\bea\label{m1RRWH}
-\log W_\sR(t_{i|R}, t_{f|R})&=&h\log\left[\frac{C^2\sinh^2\left(\frac{\pi   \Delta t}{\beta }\right) \left(4 (C-3) \sinh^2\left(\frac{\pi  \Delta t}{\beta }\right) -9\right)}{4 \pi ^2 \mathcal{L}^2(C-3)^2 (4 C-3)\epsilon^4}\right]\,,\\\label{m1LLWH}
-\log W_\sR(t_{i|L}, t_{f|L})&=&h\log\left[\frac{ 4 \pi ^2 \mathcal{L}^2 \sinh ^2\left(\frac{\pi   \Delta t}{\beta }\right) }{C^2(C-3)^2 (4 C-3)\epsilon^4}\right]\quad+\\\nonumber
\\
\nonumber
&&h\log\left[(4 (C-3) ((C-6) C+4)^2 \sinh ^2\left(\frac{\pi   \Delta t}{\beta }\right)-(5 (C-4) C+12)^2\right]\,,\,
\\
\nonumber\\
\label{m1RLWH}
-\log W_\sR(t_{i|R}, t_{f|L})&=&h\log \left[\frac{4 (C-3) ((C-6) C+4) \cosh^4\left(\frac{\pi  \Delta t}{\beta }\right)}{(C-3)^2 (4 C-3)\epsilon^4}\quad+\right.\\
&&\qquad\qquad\left.\frac{(C (9 C-38)+24) \cosh^2\left(\frac{\pi \Delta t}{\beta }\right)+4 C-3}{(C-3)^2 (4 C-3)\epsilon^4}\right]\,,\nonumber
\eea
where $C$ is given in \eqref{constraintsC}, and we used \eqref{Wilsonm1m2t}.  When both endpoints are at the $R$ (or $L$) boundary we have $\Delta t= t_i-t_f$; when the endpoints are at different boundaries we have instead $\Delta t= t_i+t_f$. We should note that this reversal of the time coordinate on the left side may seem artificial, as in this gauge there is no notion of the bulk degeneration of the Killing direction; we perform it here largely for consistency with later sections, where it follows naturally. These expressions are valid for finite $(\mu,\beta)$ (or alternative finite charges $({\cal L},{\cal W})$). 

From \eqref{m1RRWH}-\eqref{m1RLWH} we can draw many conclusions about the causal properties of the wormhole gauge. First, the solution is not symmetric with respect to the two boundaries $R$ and $L$:\footnote{This asymmetry is not an artefact of the position of the boundaries: the answers cannot be made symmetric by a  rescaling of the cutoff $\epsilon$ at each boundary.} $W_\sR(t_{i|R}, t_{f|R})\neq W_\sR(-t_{i|L}, -t_{f|L})$. This already violates one of the equalities listed in \eqref{eq:kmscorr}.  Second, it is evident as well that $W_\sR(t_{i|R}, t_{f|R} )\neq W_\sR(-t_{i|R}- i\beta/2, t_{f|L})$: the wormhole gauge does {\it not} satisfy the last equality in \eqref{eq:kmscorr}. This solution cannot be interpreted as thermofield state.   

\begin{figure}
\begin{center}
\includegraphics[width=0.45\textwidth,page=1]{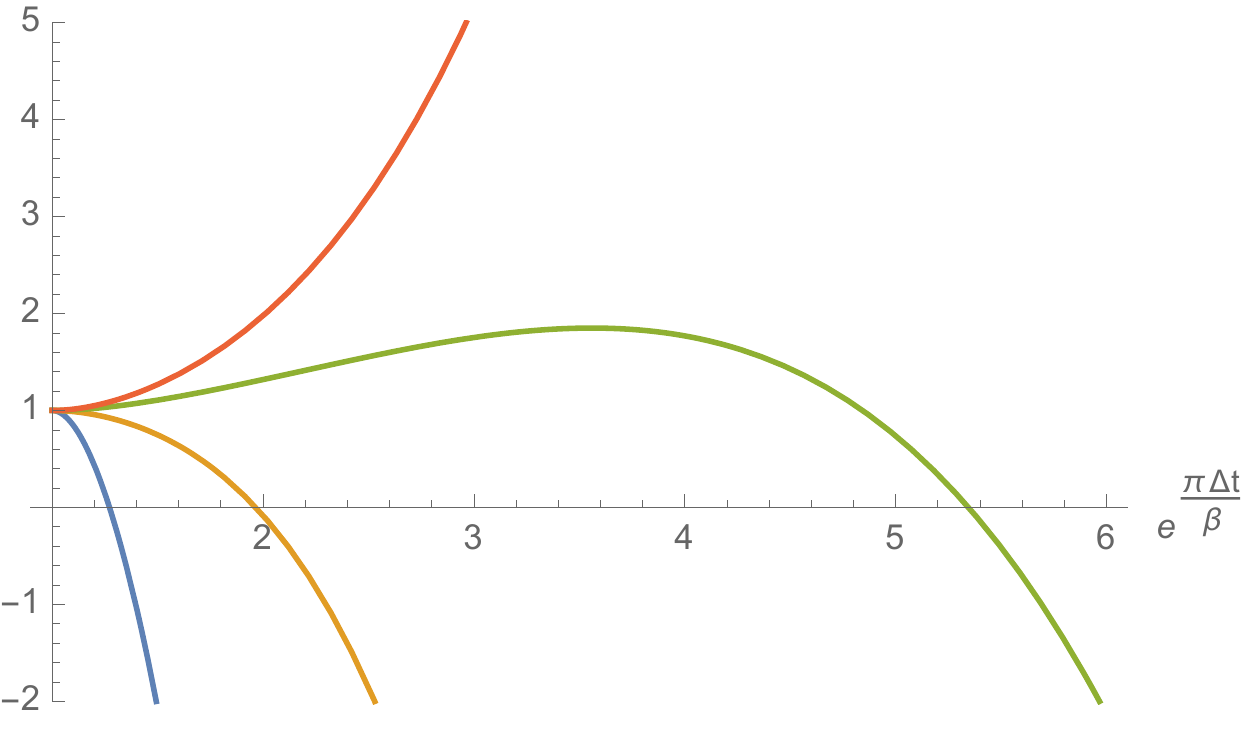}\qquad\includegraphics[width=0.45\textwidth,page=1]{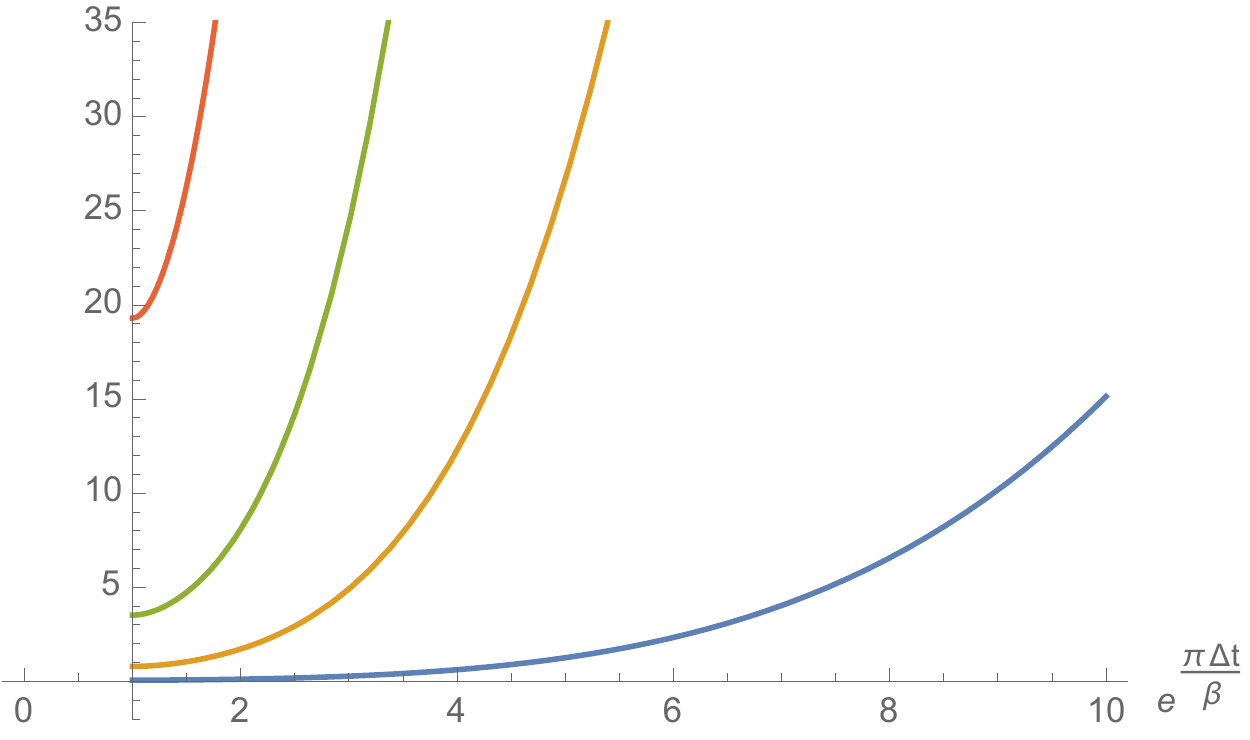}
\end{center}
\caption{Plot for the universal part of $W^{-1}_\sR(t_{i|R}, t_{f|L})$ in the wormhole gauge (left), and horizon gauge (right). We set $h=\mathcal{L}=1$ and the cutoffs are set to one (i.e. only the universal piece is plotted). The different curves correspond to different values of $C$: $C=3.3$ (blue), $C=4$ (yellow), $C=5$ (green), and $C=7.5$ (red). We see that for the wormhole gauge the correlator can change sign,  while for the horizon gauge it is always positive.}\label{fig:m1positive}
\end{figure}

 Related to the two above properties, a third feature is as follows: the argument in the logarithm of \eqref{m1RLWH} has a zero at 
 \be\label{zerom1tWH}
\cosh^2\left(\frac{\pi\Delta t}{\beta}\right)={\frac{-24+C \left(38-9 C-\sqrt{C (17 C-60)+36}\right)}{8(C-3) ((C-6) C+4)}}~,
 \ee
which has a real solution for $\Delta t$ when $\sqrt{5}+3>C>3$. This illustrates that a two-sided correlator will change sign depending on the  time separation for this range of $C$;  see  Figure \ref{fig:m1positive}. If we interpret \eqref{m1RLWH} as a geodesic distance between the two boundaries, it means that the separation between $L$ and $R$ can be either timelike, null or spacelike depending on $\Delta t$.  Hence we can send timelike signals between the two sides in the wormwhole gauge, which obviously does not fit the causal properties we would attribute to an eternal black hole.

It is instructive to compare our analysis with the one performed in \cite{Kraus:2012uf}. There  they evaluated two-sided correlators in a first order expansion about $\mu\rightarrow 0$ for the scalar field in Vasiliev theory. This field has a non-zero spin-$3$ charge: we may mimic their analysis by considering a Wilson line with non-vanishing $w_3$ to find 
\begin{align}\label{Wilsons3RRWH}
W_\sR(x_{i|R}, x_{f|R})=\frac{4 w_3\pi\mu}{3\beta}\frac{-3\sinh \left(\frac{2\pi   (\Delta \phi+\Delta t)}{\beta }\right)+\frac{2\sigma \pi\Delta \phi}{\beta}\left( \cosh \left(\frac{2\pi(\Delta \phi+\Delta t)}{\beta }\right)+2\right)}{\sinh ^2\left(\frac{\pi   (\Delta \phi+\Delta t)}{\beta }\right)}+(\Delta t\leftrightarrow -\Delta t) + \ldots\,,
\\\label{Wilsons3RLWH}
W_\sR(x_{i|R}, x_{f|L})=\frac{4 w_3 \pi\mu}{3\beta}\, \frac{\sinh \left(\frac{2\pi   (\Delta \phi+\Delta t)}{\beta }\right)+\frac{2 \sigma \pi\Delta \phi}{\beta}\left( \cosh \left(\frac{2\pi   (\Delta \phi+\Delta t)}{\beta }\right)-2\right)}{\cosh ^2\left(\frac{\pi   (\Delta \phi+\Delta t)}{\beta }\right)}+(\Delta t\leftrightarrow -\Delta t)+ \ldots \,,\quad\,
\end{align}
where we are only displaying the linear term in $\mu$-expansion of the Wilson line. The result above is in perfect agreement with the expression in \cite{Kraus:2012uf,Gaberdiel:2013jca}. The first order correction \eqref{Wilsons3RLWH} does not have a singularity, and this suggests that the two boundaries are causally disconnected as argued in \cite{Kraus:2012uf}. However, as illustrated by \eqref{zerom1tWH}, this apparent regularity is an artifact of the $\mu$ expansion: over a finite range of $C$ the correlator allows for timelike geodesics. 

Based on this analysis, we would attribute to the wormhole gauge a Penrose diagram with a rectangular shape where signals can cross from one boundary to another. Even though this solution has no thermofield double interpretation, we should keep in mind that the result for the $R$ side correlators are compatible with CFT computations. This agreement with the dual theory is an important feature to preserve as we build the connections associated with the thermofield state. 

\subsection{Failures and successes of the horizon gauge}\label{sec:uc}


We could attribute the failure of the wormwhole gauge to the lack of a preferred point in the geometry that we can associate with a horizon. The first attempt to fix this feature was discussed in \cite{Ammon:2011nk}. They considered connections for which the radial function in \eqref{eq:aba} is modified as follows
\begin{align}\label{BHgaugegb}
 A(r)=b(r)^{-1}(a+d)\,b(r)\,,\qquad\qquad b(r)= e^{(r+\rho_0)L_0}g(r)\,,\\ \nonumber
  \bar A(r)=\bar b(r)\,(\bar{a}+d)\bar b(r)^{-1}\,,\qquad\qquad \bar b(r)=g(r) e^{(r+\rho_0)L_0}\,.
\end{align}
At this stage $\rho_0$ is a free parameter, which in \cite{Ammon:2011nk} is set to be equal to the BTZ value\eqref{magic}. The group element $g(r)$ is fixed by demanding that the connections satisfy 
\begin{align}
 & A_t(-r)=h^{-1}(r)\bar A_t(r) h(r)\,,\label{BHgauget}\\
 & A_\phi(-r)=-h^{-1}(r)\bar A_\phi(r) h(r)\,,\label{BHgaugex}
\end{align}
with $h(r)\in SL(3,\,\mathbb{R})$ and arbitrary modulo the condition $h(0)=\mathds{1}$. In \cite{Ammon:2011nk}, one explicit combination of $h(r)$ and $g(r)$ is found that fulfils the above conditions. These results are reviewed in the Appendix \ref{app:gauge}. This construction provides a smooth horizon for the static holomorphic and canonical black hole. The motivation is quite natural: it is a generalization of the condition that the time component of the generalized vielbein $A_t(r) - \bA_t(r)$ vanishes at a point. Their construction assures smoothness of the metric and spin-3 field around the horizon at $r=0$, and for this reason we denote this construction as {\it horizon gauge}. 

The horizon gauge is compatible with weak Kruskal gauge defined in Section \ref{sec:refeuc}. Both conditions imply the vanishing of $A_t(r)-\bar A_t(r)$ at the origin, and moreover we have verified that the combination $R(y)L(y)$ is a smooth function at the origin of the Euclidean disc. The real difference lies not at the horizon but at infinity: essentially the relations imposed above between $A$ and $\bar{A}$ at all values of $r$ seem to fix the behavior of $b(r), \bb(r)$ everywhere. In particular, they do not approach the usual asymptotically AdS choice $b(r) \sim e^{r L_0}$ at infinity: this means that the CFT interpretation of this gauge choice is obscure, and has implications for correlation functions as computed using the Wilson line. 


As in the wormhole case, we would like to analyze the features of the Wilson line for the horizon gauge. We consider first the case $\Delta \phi=0$, and we compute the leading order of the Wilson line in the cutoff, which is denoted by $\varepsilon$ in this case.  Using the results of Appendix \ref{app:gauge} and \ref{app:wilson}, we obtain
\begin{align}\nonumber
-\log W_\sR(t_{i|R}, t_{f|R})&=
-\log W_\sR(t_{i|L}, t_{f|L}) \\ &=h\log \left(\frac{3 \beta   \sinh \left(\frac{\pi  \Delta t}{\beta }\right)}{8 \pi  \mu \varepsilon}\right)^4-\frac{16 h \pi ^2 \mu ^2 \left( 31+\text{csch}^2\left(\frac{\pi \Delta t}{\beta }\right)\right)}{9 \beta ^2}+ O(\mu^4)\,,  \label{m1RRBH}
\\\label{m1RLBH}
-\log W_\sR(t_{i|R}, t_{f|L})&=h\log \left(\frac{3 \beta   \cosh \left(\frac{\pi \Delta t}{\beta }\right)}{8 \pi  \mu \varepsilon }\right)^4-\frac{16 h \pi ^2 \mu ^2 \left( 31-\text{sech}^2\left(\frac{\pi  \Delta t}{\beta }\right)\right)}{9 \beta ^2}+ O(\mu^4)\,.
\end{align}
The full expression for the time correlators in the black hole gauge is less gentle to the eye than for the wormhole, and for this reason we only show the two first terms in the expansion around $\mu\rightarrow 0$. We see from \eqref{m1RRBH}-\eqref{m1RLBH}  that at leading order in $\mu$ the KMS conditions in \eqref{eq:kmscorr} hold; this persists at all orders in the $\mu$-expansion. 
Therefore, the correlation functions of the blackhole gauge do have the features of a thermofield double state.

To analyze if the two sides are connected or disconnected, analogously as we did in \ref{sec:WH}, we should consider all terms in the $\mu$-expansion of $W_\sR(t_{i|R}, t_{f|L})$.  Since the expression is more cumbersome for finite $\mu$,  we plotted $W_\sR(t_{i|R}, t_{f|L})$ for a wide range of values $C$, and found that it is always positive (see Fig. \ref{fig:m1positive}). This is in complete agreement that the horizon gauge has two causally disconnected sides, as it should.

However, there are some problems when we compare these answers with the results from wormhole gauge (which itself agrees with the CFT, as described earlier). For instance, if we  expand the wormhole solution \eqref{m1RRWH} around $\mu\rightarrow 0$ the result is
\be
\label{m1RRBHmu}
-\log W_\sR(t_{i|R}, t_{f|R})= h\, \log \left(\frac{\beta ^4\sinh ^4\left(\frac{\pi\Delta t}{\beta }\right)}{\pi ^4 \epsilon ^4}\right)-\frac{16 h \mu ^2 \pi ^2 \left(3\,\text{csch}^2\left(\frac{\pi  \Delta t}{\beta }\right)+5\right)}{3 \beta ^2}+ O(\mu^4)\,.
\ee
It is evident that  \eqref{m1RRBH} is not equal to \eqref{m1RRBHmu} even if we try to adjust the cutoff $\epsilon$ and $\varepsilon$. A similar problem occurs if we consider spatial separations. Thus the horizon gauge does not reproduce the known results of two point functions for spin-3 operators in a CFT with ${\cal W}_3$ symmetry.

\subsection{A successful gauge}

In this last portion we report on the values of the Wilson line for the strong Kruskal gauge. As discussed in section \ref{sec:defBH}, this gauge is defined by demanding that $L(y)$ and $R(y)$ are smooth functions near the Euclidean origin. This imposes restrictions on the radial functions $b(r)$ and $\bb(r)$; in Appendix \ref{app:kruskal} we demonstrate how to build a solution to these regularity conditions while preserving the asymptotic behavior. Note that once we know that a solution exists, we do not actually need to use its explicit form to calculate correlators: since we are imposing AdS asymptotics at the $R$ boundary, one-sided correlators will agree with those computed from the wormhole gauge above. Furthermore by design of the strong Kruskal gauge, the Wilson line that interpolates between $L$ and $R$ is related to the single-sided correlator via the expected half-shift in $\beta$. 

Thus for equal space separation the values of $W_{\sR}(C_{ij})$ are 
\bea
\log W_\sR(t_{i|R}, t_{f|R})&=&\log W_\sR(t_{i|L}, t_{f|L}) \nonumber\\ \label{m1RRLLf}
&=&-h\log\frac{C^2\sinh^2\left(\frac{\pi   \Delta t}{\beta }\right) \left(4 (C-3) \sinh^2\left(\frac{\pi  \Delta t}{\beta }\right) -9\right)}{4 \pi ^2 \mathcal{L}^2(C-3)^2 (4 C-3)\epsilon^4}\,,\\\label{m1RLf}
\log W_\sR(t_{i|R}, t_{f|L})&=&-h\log\frac{C^2\cosh^2\left(\frac{\pi   \Delta t}{\beta }\right) \left(4 (C-3) \cosh^2\left(\frac{\pi  \Delta t}{\beta }\right) +9\right)}{4 \pi ^2 \mathcal{L}^2(C-3)^2 (4 C-3)\epsilon^4}\,.
\eea
Recall that these expressions hold for both holomorphic and canonical black holes since the time component of $(A, \bar A)$ is the same.  
Since $C\geq 3$, the argument of logarithm of \eqref{m1RLf} is always positive: at zero spatial separation  the two sides are causally disconnected for all ranges of $\Delta t$. This gauge is so far compatible with the expected properties of an eternal higher spin black hole, and it reproduces correctly the known results in the CFT (which involve setting $\Delta t=0$ and a fix charge configuration without chemical potentials).

\begin{figure}
\begin{center}
\includegraphics[width=0.5\textwidth]{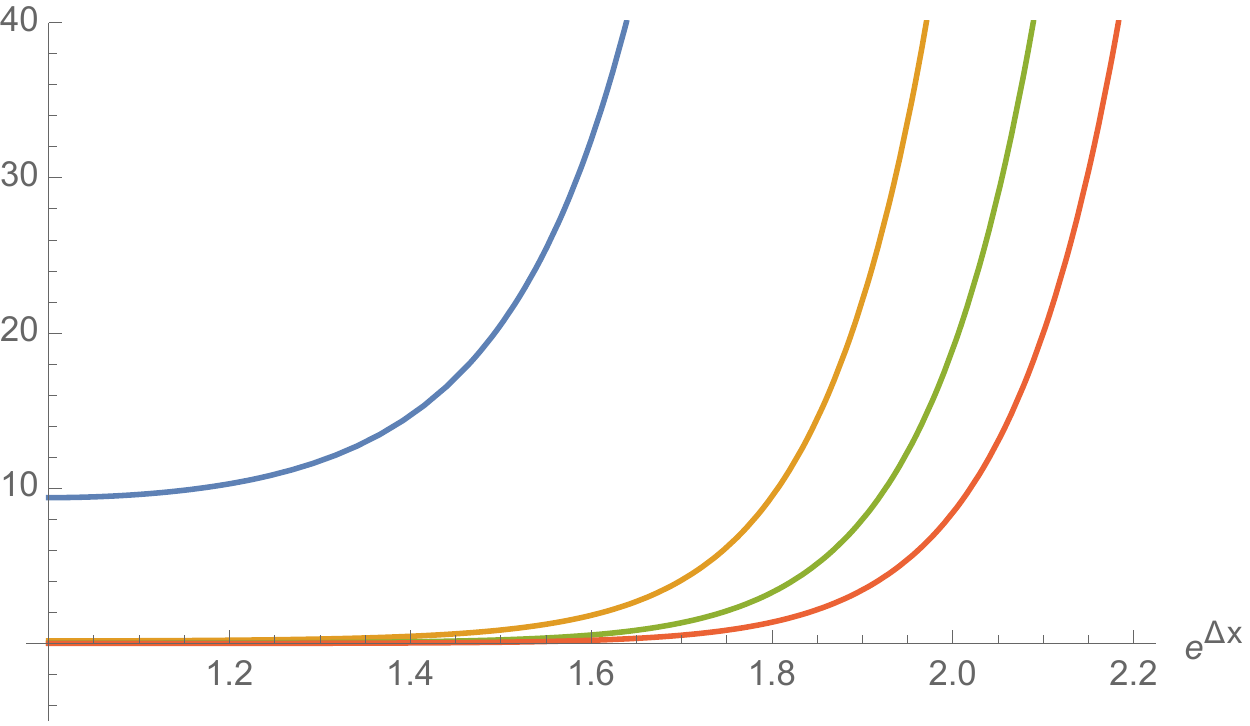}\includegraphics[width=0.5\textwidth]{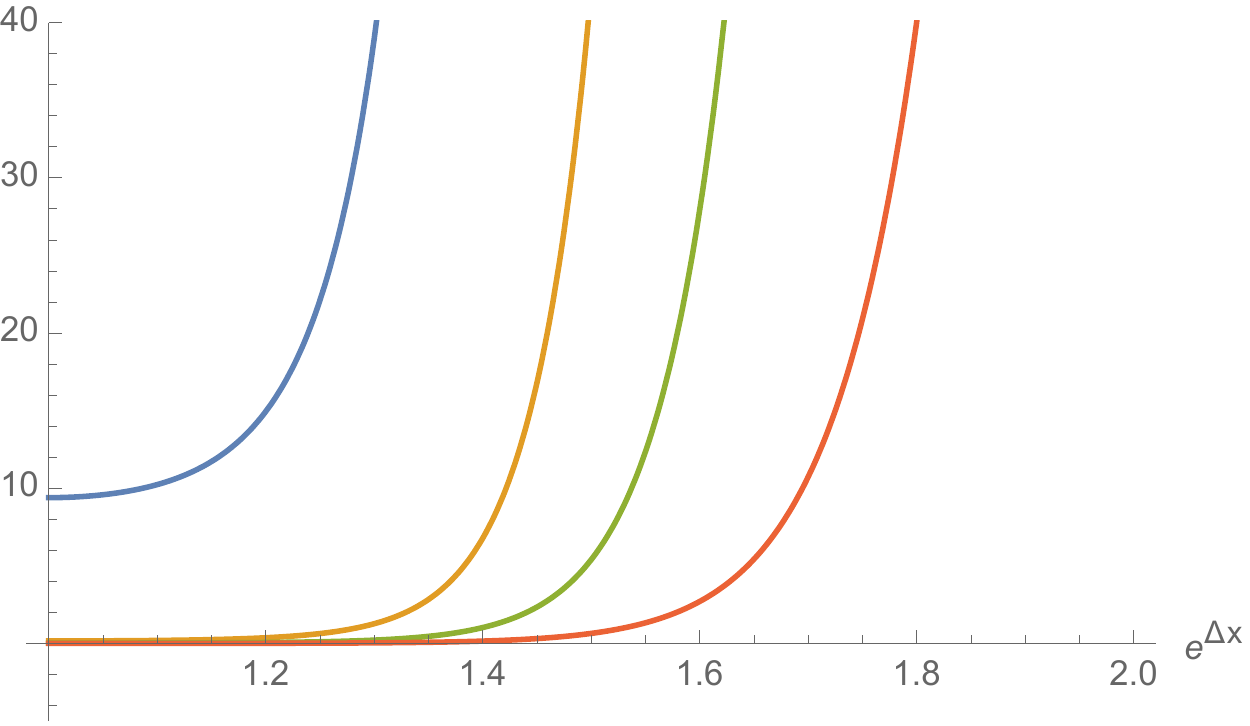}
\end{center}
\caption{Plot of the universal part of $W^{-1}_\sR(x_{i|R}, x_{f|L})$ when $h=2$, $w_3=0$ for canonical (left), and holomorphic black hole (right). Here $\mathcal{L}=1$ and we removed the cutoff.  The different curves correspond to different values of $C$: $C=3.3$ (blue), $C=4$ (yellow), $C=5$ (green), and $C=7.5$ (red). We see that for both black holes the correlator is always positive.}\label{fig:RLsucc}
\end{figure}

It is also useful to record the values of the Wilson lines for $\Delta t =0$. If the probe is not charged, i.e. $w_3=0$, we have
\bea
\log W_\sR(x_{i|R}, x_{f|R}) &=& - h\log\left[
\frac{ \left(1+\frac{3}{\sqrt{4 C-3}}\right) e^{\lambda_1\Delta \phi}-2\, e^{\lambda_2\Delta \phi}+\left(1-\frac{3}{\sqrt{4 C-3}}\right) e^{\lambda_3\Delta \phi}}{8\pi  \mathcal{L} (C-3)C^{-1}\epsilon^2}\times\right.\nonumber\\
&&\left.\frac{ \left(1+\frac{3}{\sqrt{4 C-3}}\right) e^{-\lambda_1\Delta \phi}-2\, e^{-\lambda_2\Delta \phi}+\left(1-\frac{3}{\sqrt{4 C-3}}\right) e^{-\lambda_3\Delta \phi}}{8 \pi \mathcal{L} (C-3) C^{-1}\epsilon^2}\right]~,
\cr
\log W_\sR(x_{i|R}, x_{f|L}) &=&   -h\log\left[
\frac{ \left(1+\frac{3}{\sqrt{4 C-3}}\right) e^{\lambda_1\Delta \phi}+2\, e^{\lambda_2\Delta \phi}+\left(1-\frac{3}{\sqrt{4 C-3}}\right) e^{\lambda_3\Delta \phi}}{8\pi  \mathcal{L} (C-3) C^{-1} \epsilon^2}\times\right.\nonumber\\
&&\left.\frac{ \left(1+\frac{3}{\sqrt{4 C-3}}\right) e^{-\lambda_1\Delta \phi}+2\, e^{-\lambda_2\Delta \phi}+\left(1-\frac{3}{\sqrt{4 C-3}}\right) e^{-\lambda_3\Delta \phi}}{8 \pi \mathcal{L} (C-3) C^{-1} \epsilon^2}\right]~. \label{m1RRx}
\eea
These functions are plotted in Figure \ref{fig:RLsucc}. Here $\lambda_i$ are the eigenvalues of $a_\phi$ component of the connection which read
\bea
 (\lambda_1,\lambda_2, \lambda_3)  = \sqrt{\frac{2 \pi \mathcal{L}}{C}}\left(-\sigma +\sqrt{4 C-3}\frac{2 C +3(\sigma -2) }{2 C-3},\, 2 \sigma,\, -\sigma -\sqrt{4 C-3} \frac{2 C+3(\sigma -2)}{2 C-3}\right)
\eea
and $\sigma$ controls if the solution is holomorphic or canonical as defined in \eqref{eq:sigma}. 

\begin{figure}
\begin{center}
\includegraphics[width=0.5\textwidth]{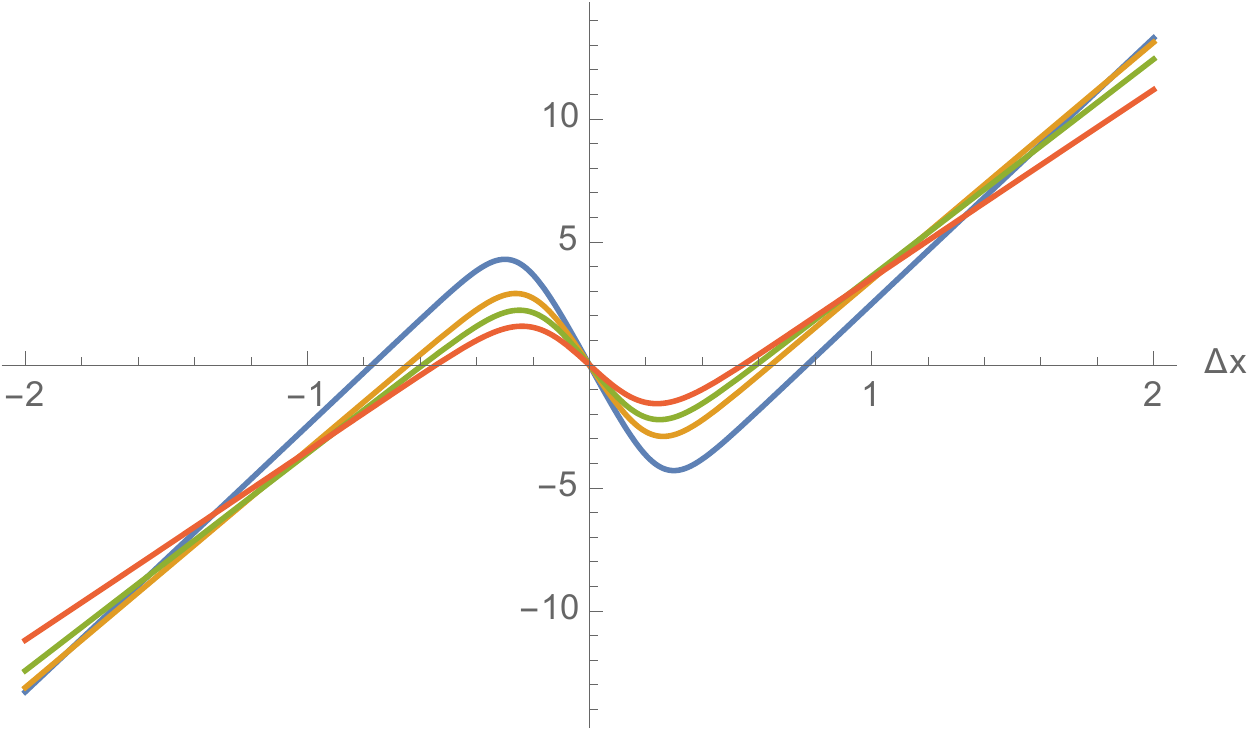}\includegraphics[width=0.5\textwidth]{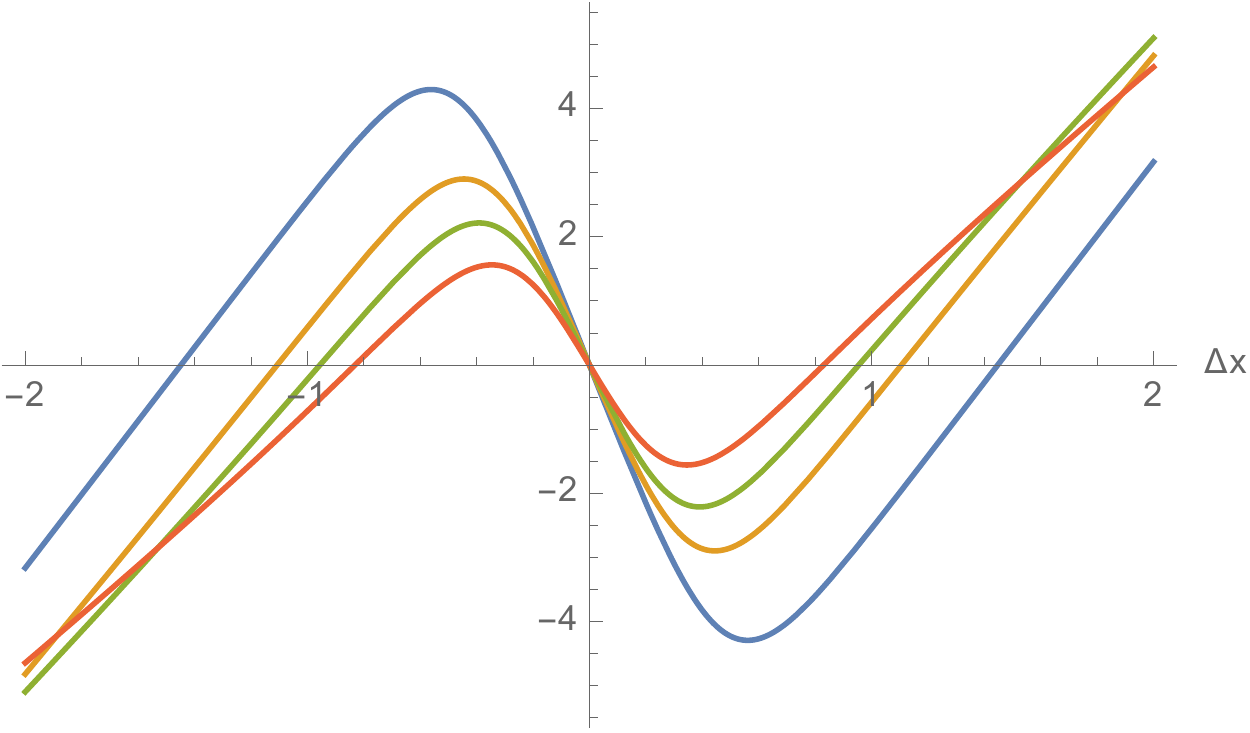}
\end{center}
\caption{Plot of  $-\log W_\sR(x_{i|R}, x_{f|L})$  when $h=0$, $w_3=1$ for  in the canonical (left), and holomorphic black hole (right).  The different curves correspond to different values of $C$: $C=3.3$ (blue), $C=4$ (yellow), $C=5$ (green), and $C=7.5$ (red).}\label{fig:RLsuccspin3}
\end{figure}

It is also interesting to evaluate the Wilson line with $h=0$ and $w_3\neq0$: this would correspond to a probe that only carries higher spin charge.  In  Figure \ref{fig:RLsuccspin3} we plot the behavior of such a Wilson line between two spatially separated points on the two boundaries. It is interesting that there is a reflection symmetry associated with flipping the sign of the spatial direction together with the higher spin charge of the probe: $ W_\sR(x_{i|R}, x_{f|L})\big|_{w_3}=W_\sR(-x_{i|R}, -x_{f|L})\big|_{-w_3}$. The behavior shown here may be interpreted as a potential well felt by the charged probe arising from its coupling to the background spin-$3$ field. It would be interesting to explore further the implications of such non-monotonic behavior. 

\section{Applications}\label{sec:appl}

In this section we explore various properties of the thermofield state in higher spin gravity as accessed by the two-sided black hole in Kruksal gauge. We perform our computations in the canonical black hole: for the most part, the results for the holomorphic black hole are very similar, except where the complications in the holomorphic black hole discussed around \eqref{eq:sigma} manifest themselves. 

\subsection{Higher spin black hole interiors and entanglement velocities}
It is well-known that the interior of a ordinary (spin-$2$) eternal black hole grows as one moves ``upwards''  in time (i.e. in the time direction $\p_{t_{L}} + \p_{t_R}$ that is orthogonal to the Killing direction). It was demonstrated in \cite{Hartman:2013qma} that this growth can be given a simple field-theoretical interpretation in terms of the time-dependence of entanglement entropy. 

\begin{figure}
\begin{center}
\includegraphics[width=0.5\textwidth,page=5]{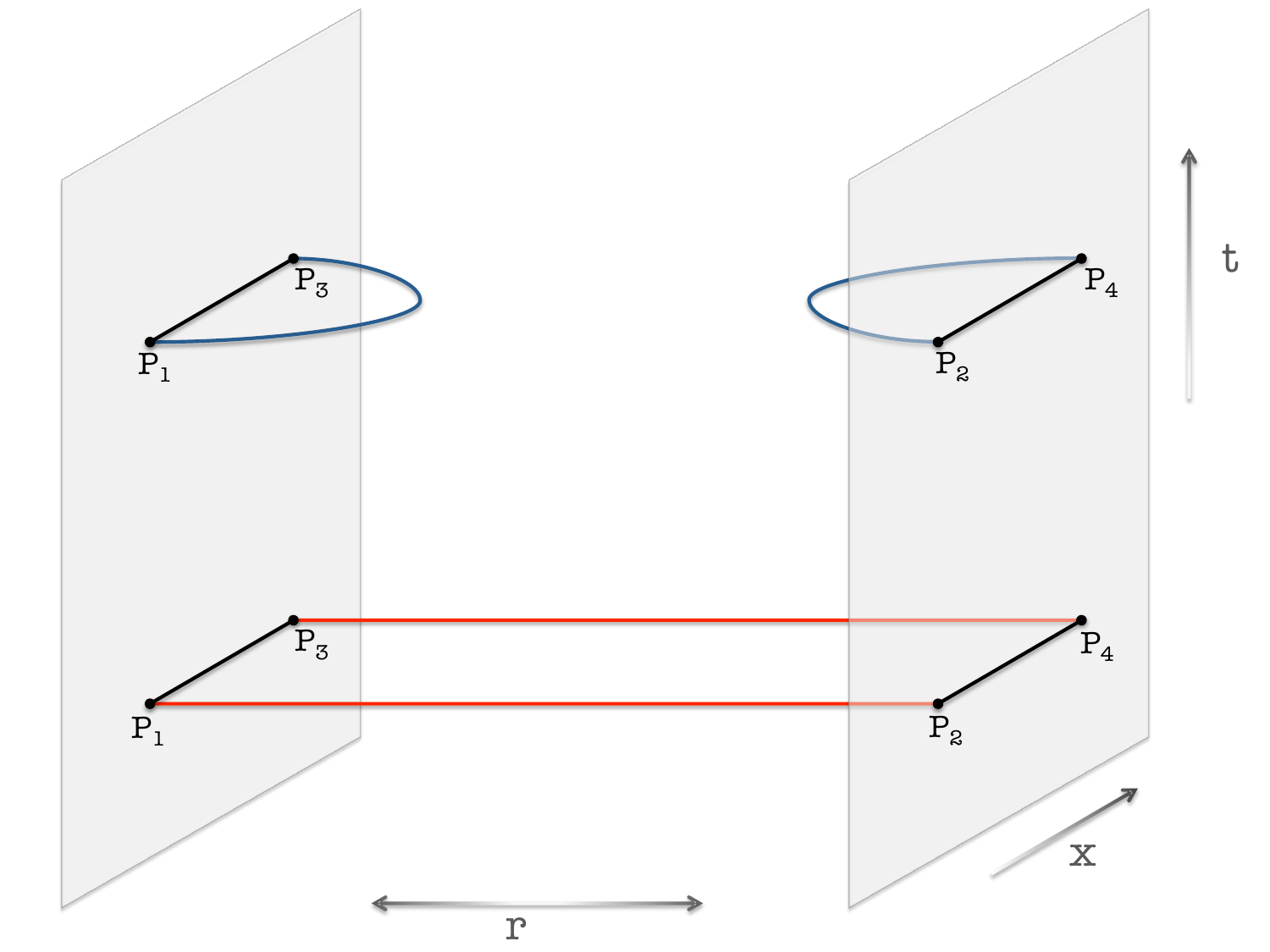}
\end{center}
\caption{Configurations for time evolution of entanglement entropy in the thermofield state. From the bulk perspective it is a competition between the connected configuration (red lines), which gives \eqref{Wilcross}, and the disconnected contribution (blue lines) in \eqref{Sdiv}.}\label{fig:scrambling}
\end{figure}

We briefly review the setup: recall from \eqref{eq:thf1} that the thermofield state is given by

\be\label{eq:thfrep}
|\psi \rangle = \frac{1}{\sqrt{Z}}\sum_n e^{-\frac{\beta}{2}H} |{\cal U} n\rangle_L \otimes |n \rangle_R~, \qquad H|n\rangle = (E_n + \mu Q_n)|n\rangle~,
\ee
where we have included the deformation by the chemical potential in the Hamiltonian. One can now define a one-parameter family of states by acting on this state with the sum of the left and right Hamiltonians:
\be
|\psi(t_{\star})\rangle \equiv e^{i(H_L + H_R)t_{\star}}|\psi\rangle \ . \label{timeevolv}
\ee
This action moves us ``upwards'' in time (note that the orthogonal action of $H_L - H_R$ leaves $|\psi\rangle$ invariant, and corresponds in the bulk to the Killing direction).  Consider now the entanglement entropy in the state $|\psi(t_\star)\rangle$ of a region given by the union of two intervals, one in the left CFT and one in the right, both of length $\Delta\phi$. This may be computed holographically by considering the geometry shown in Figure \ref{fig:scrambling}, where the endpoints on each side are separated by a distance $\Delta\phi$ and are located at $t_L = t_R = t_{\star}$, where $t_{\star}$ increases as we move upwards. 

In our setup, there are two configurations of Wilson lines that contribute: one set of Wilson lines joins each endpoint of an interval with its partner in the other CFT by crossing through the black hole interior. Its contribution can be found from \eqref{m1RRLLf} with $\Delta t = t_L + t_R = 2t_{\star}$ to be
\bea\label{Wilcross}
S_{conn} = -2\log W_\sR(t_{i|R}, t_{f|L})&\sim& 2 h\log\frac{ C^2   e^{\frac{8\pi t_{\star}}{\beta }}}{16 \pi ^2 \mathcal{L}^2(C-3) (4 C-3)\epsilon^4}\cr &=&\frac{16h\pi  t_{\star}}{\beta }+S_{div}~,
\eea
where we have specialized to times $t_{\star} \gg \beta$ and where
\be\label{Sdiv}
S_{div}\equiv 2h\log\left[\frac{ C^2 }{16 \pi ^2 \mathcal{L}^2(C-3) (4 C-3)\epsilon^4}\right]~.
\ee
The second configuration contains two Wilson lines that each remain outside the black hole horizon.  In this case, the result will be given by twice the one-sided Wilson line in \eqref{m1RRx} with  $\Delta t=0$. We will consider the limit $\Delta \phi\gg \beta$: extracting the dominant long-distance contribution we find 
\bea\label{Wiloneside}
S_{disc} = -2\log W_\sR(x_{i|R}, x_{f|R})&\sim& 2 h\log\frac{ C^2   e^{(\lambda_1-\lambda_3)\Delta\phi}}{16 \pi ^2 \mathcal{L}^2(C-3) (4 C-3)\epsilon^4} \cr &=& 4h \sqrt{\frac{2\pi \sL(4C -3)}{C}} \Delta\phi+S_{div}~.
\eea
 Up till now we have focused on Wilson lines as computing two-point functions of light operators. However, as was argued in \cite{deBoer:2013vca,Ammon:2013hba}, these Wilson lines also compute entanglement entropy if one evaluates them at the precise dimension $h \to \frac{c}{12}$. It is convenient to write the above results in terms of the entropy density in units of the inverse temperature:
\be
s = \frac{c}{6}\le(\frac{\pi}{\beta}\frac{2C - 3}{C-3}\ri)~.
\ee 
We now normalize the results with this entropy density to find:
\be 
S_{conn} = 4 s v t_{\star} + S_{div}~, \qquad S_{disc} = 2 s \Delta\phi + S_{div}~,
\ee
where for the time-dependent configuration we have defined an entanglement velocity
\be
v \equiv \frac{C-3}{C - \frac{3}{2}} ~ ,
\ee 
which interpolates from $v \to 1$ at zero higher spin charge to $v \to 0$ as we approach the extremal higher spin black hole. We see that at small times the answer is dominated by the connected configuration; however as time goes on the inside of the black hole grows linearly in size, and the connected configuration becomes energetically more and more expensive. Eventually there is a phase transition to the disconnected configuration at $t_{\star} = \frac{\Delta\phi}{2 v}$, after which the entanglement entropy saturates at its thermal value. 

The interpretation of these results is standard \cite{Calabrese:2005in,Hartman:2013qma}. The time evolution of entanglement entropy in $1+1$ dimensional systems may be viewed in terms of a quasi-particle picture: if the initial state is excited but has essentially only short-range entanglement, then upon time evolution the entanglement entropy grows as entangled pairs of particles stream across the endpoints of the interval, entangling the interior with the outside. The two-sided time evolution in \eqref{timeevolv} fits into this picture with the slight modification that we now consider entanglement across the two CFTs \cite{Hartman:2013qma}.  The entanglement velocity $v$ defined above then quantifies how quickly these quasiparticles move: apparently as we approach extremality the entangling particles slow down to zero speed, perhaps due to scattering off of the large density of higher spin charge present. A similar result for the entangling velocity as a function of chemical potential has been derived in the context of Reissner-Nordstrom black holes in higher dimensions \cite{Liu:2013iza,Liu:2013qca}. It would be interesting to further understand the dependence of the velocity on the background charge density from a field-theoretical point of view.  

To summarize: as probed by the entanglement entropy, the interior of a two-sided higher spin black hole grows with time, as expected from basic field theoretical notions of the time evolution of entanglement entropy. There are other time dependent observables worth understanding on Lorentzian higher spin backgrounds, in particular those recently reported in \cite{Roberts:2014ifa,Maldacena:2015waa,Perlmutter:2016pkf}.

We could also probe the two-sided higher spin black hole with the ``spin-3 entanglement entropy'' $S^{(3)}$ of \cite{Hijano:2014sqa}. In this context this corresponds to a probe with $h=0$, and $w_3\neq 0$ in \eqref{eq:wr3}, and taking again the arrangement of intervals in Figure \eqref{fig:scrambling}. Interestingly, however, now the configuration that interpolates between the two boundaries is trivial: $S^{(3)}_{conn}=0$. As explained in appendix \ref{app:wilson}, this is a simple consequence of the algebraic properties of the Wilson line for $\Delta \phi=0$. More generally, there is no exponential in time behavior of two point functions of this class of higher spin correlators. 

%
%
In the limit $\Delta \phi\gg \beta$ the contribution of the blue Wilson lines in Figure \eqref{fig:scrambling} is
\be\label{Wiloneside3}
S^{(3)}_{disc} = -2\log W_\sR(x_{i|R}, x_{f|R})\sim 2 w_3\log\frac{\left(\sqrt{4 C-3}+3\right) e^{( \lambda_1+\lambda_3)\Delta \phi}}{\sqrt{4 C-3}-3} =- 4w_3 \sqrt{\frac{2 \pi\mathcal{L}}{C}} \Delta\phi+S^{(3)}_{div}~.
\ee
 and we define
\be\label{Sdiv3}
S^{(3)}_{div}\equiv 2 w_3\log\frac{\sqrt{4 C-3}+3}{\sqrt{4 C-3}-3}~.
\ee
where there is no short range ``entanglement'':  $S^{(3)}$ has no UV divergent pieces. Note that in the regime of interest $\Delta\phi \gg \beta$ this one-sided contribution might be expected to never dominate the answer, as the two-sided contribution does not grow with time as it did in the conventional entanglement calculation above. It would be interesting to have a better interpretation of these higher-spin correlation functions on the thermofield state.

\subsection{Extremal black holes and an emergent AdS$_2$}
We now turn to the zero temperature limit of the higher spin black hole. It is well-known that charged black holes in higher dimensions generally develop an AdS$_2$ factor when cooled down to zero temperature. The AdS$_2$ indicates an emergent conformal symmetry at low energies that acts only on the time coordinate: this manifests itself in field-theory correlation functions, which now exhibit power-law correlations in time but have a finite correlation length in space \cite{Iqbal:2011in,Faulkner:2011tm,Faulkner:2009wj}. 

It is not clear whether an AdS$_2$ factor appears in the extremal limit for higher spin gravity. However, it is rather straight forward to take the zero-temperature limit of the correlation functions computed above. From \eqref{constraintsC} we take $\frac{\beta}{\mu} \to \infty$, while holding $\mu$ fixed,  by sending $C \to 3$ and holding ${\cal L}$ fixed. Thus from \eqref{m1RRWH} we may simply reduce $\sinh\le(\frac{\Delta t}{\beta}\ri) \to \frac{\Delta t}{\beta}$ to find:
\be
W_{\sR}(t_{i|R}, t_{f|R}) \sim  \le(\frac{\Delta t}{\mu}\ri)^{-2h}~,
\ee
up to a overall constant. This implies an emergent scale-invariance in the time direction, where the IR scaling dimension is equal to the UV dimension (i.e. $h$). Actually this power law behavior is guaranteed from the definition of extremality in \cite{Banados:2015tft}:  in a nutshell, an extremal black hole is characterized by $a_\phi$ being non-diagonalizable. Using the fact that $a_t$ and $a_{\phi}$ commute and that $a_t$ is of the form \eqref{diagA} even in the extremal limit, it is straightforward to show that at extremality $a_t$ is actually a nilpotent matrix. This means that that the exponentials of the form $e^{a_t \Delta t}$ appearing in $M$ (as defined in \eqref{eq:M1}) truncate after only a few terms, and thus that the correlators have only polynomial (and not exponential) dependence on $\Delta t$.   

On the other hand, the spatial correlation function remains non-trivial as the temperature vanishes; at large spatial separations we find
\be
W_{\sR}(x_{i|R}, x_{f|R}) \sim \exp\le(-\frac{h\sqrt{3}}{2}\frac{\Delta\phi}{\mu}\ri),
\ee 
indicating a nonzero spatial correlation length scaling with $\mu$. This is precisely the behavior mentioned above: interpreted geometrically, it suggests an AdS$_2 \times \mathbb{R}$ factorization of the higher spin geometry \cite{Iqbal:2011in}. We also note that the two-sided correlation function across the two sides \eqref{m1RLf} vanishes as $\le(\frac{\beta}{\mu}\right)^{-2h}$ we take the $\frac{\beta}{\mu} \to \infty$ limit, as one would expect from the infinite ``geodesic'' distance down an AdS$_2$ throat. 

It would be very interesting to understand if there is indeed an emergent $\slt$ acting on the bulk gauge connections in the extremal limit, perhaps following the algebraic approach of \cite{Banados:2015tft}. For this one would need a notion of `near horizon geometry' in Chern-Simons theory, and within this region to argue that there is an enhancement of the symmetries of the extremal solution.

\section{Discussion} \label{sec:disc}

In this work we motivated and implemented a definition of eternal black holes in the Chern-Simons formulation of higher spin gravity. Our definition introduces the concept of {\it strong (weak) Kruskal gauge} as explained in Section  \ref{sec:defBH}. A key ingredient to test our definition was the evaluation of the Wilson line defined in \cite{Ammon:2013hba,deBoer:2013vca}. This object was used as a probe of causality of a given Lorentzian background: it is the natural replacement of geodesic distances in higher spin gravity. The basic configurations we considered are presented in Figure \ref{fig:lor1}. 
 \begin{figure}
\begin{center}
\includegraphics[width=0.3\textwidth,page=1]{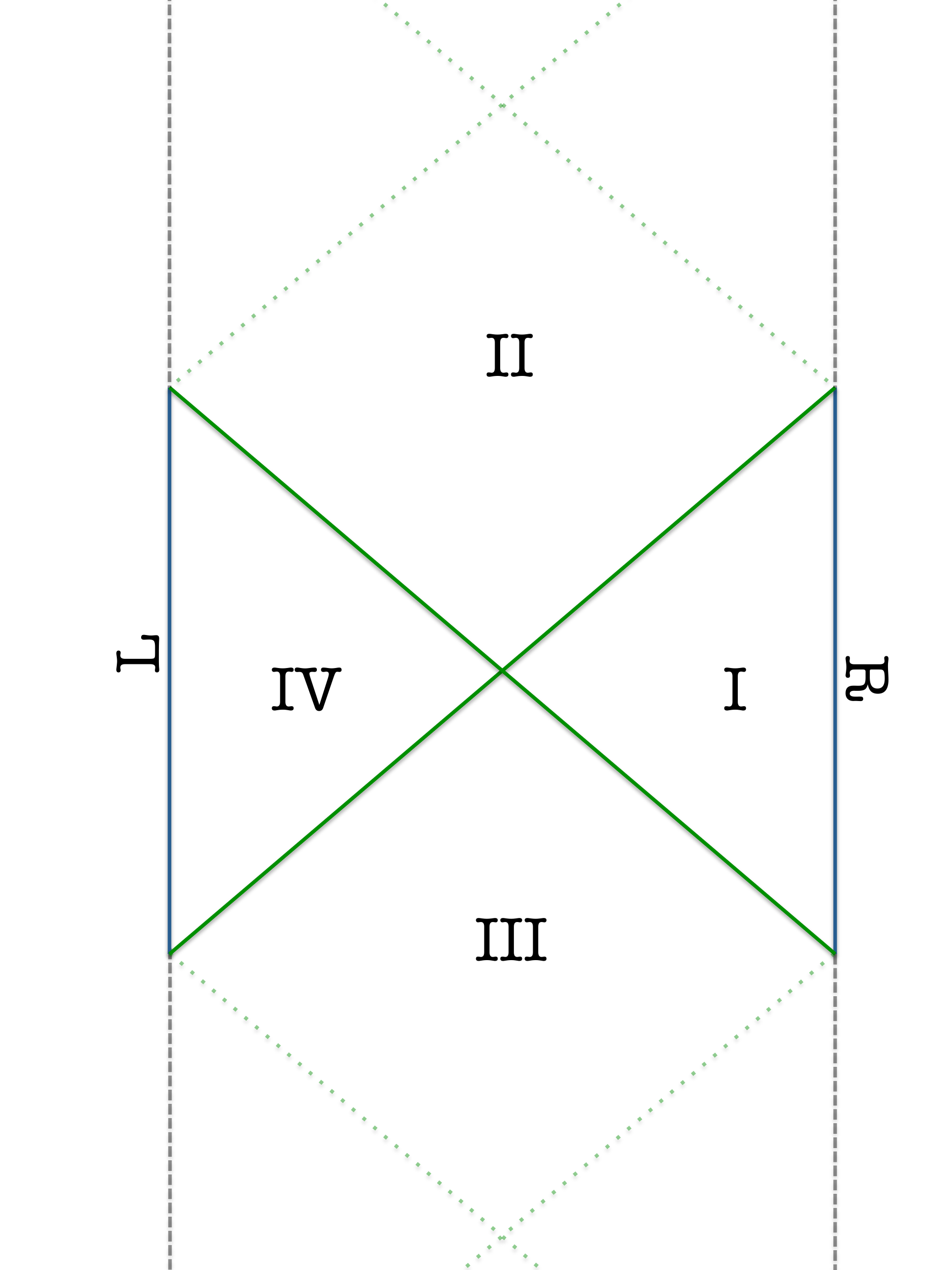}
\end{center}
\caption{Penrose Diagram for an $sl(3)$ black hole. The green line correspond to the past and future outer horizon. The dotted and dashed line would tentatively be the locations of inner horizons and singularities.}\label{fig:hsp}
\end{figure}

Our proposal was tested in a variety of ways with the two most salient points being
\begin{enumerate}
\item In the Chern-Simons formulation of $\slt$ gravity, we showed how our prescription permits access to the entire maximally extended spacetime for static (non-rotating) configurations. This illustrates explicitly that our refined definition of regularity agrees with the Lorentzian definitions in metric like formulation of gravity. 
\item On general grounds, it is expected that an eternal black hole behaves in the dual theory as the thermofield state. Two point functions on this state satisfy KMS conditions \eqref{eq:kmscorr}. Our definition exactly reproduces these conditions. 
\end{enumerate}
Using this definition we built explicitly the strong Kruskal gauge for a higher spin black hole in $sl(3)$ gravity. The tentative Penrose diagram that we would attribute to this solution is shown in Figure \ref{fig:hsp}. What our analysis clearly establishes are the existence of regions {\bf I} and {\bf IV} in the Penrose diagram. However there are some further properties of this diagram that remain puzzling. In particular, some concepts that are not addressed here include
\begin{enumerate}
\item {\it Singularity:} Due to the topological nature of the three dimensional gravitational theories we are studying, there are no curvature singularities. The line denoted ``singularity''  in the Penrose diagram for BTZ (see Figure \ref{fig:btz}) refers to a pathology of the quotient: after the dotted line there are closed timelike curves \cite{Banados:1992gq}. It is not clear to us what is the description of this singularity in Chern-Simons formulation, and hence its generalization to higher spin gravity remains an open question.
\item {\it Inner horizons:} The $sl(3)$ black hole we studied here has two free parameters: its mass $\cal L$ and spin-3 charge $\cal W$. Hence it is natural to speculate that its global properties should mimic those of a Reissner-Nordstrom solution in four dimensions. In particular, since the   $sl(3)$ black hole has a non-trivial extremal limit, there should be a notion of inner horizon and the extremal case would correspond to the confluence of these horizons. However, this is another definition that is not clear how to capture in Chern-Simons theory. One reason this is puzzling is the following: what is the size of the inner horizon of a black hole in Chern-Simons theory? There is no need to consider higher spin gravity, since this question can be phrased for the rotating BTZ black hole. By design, the holonomy of the connections along $\phi$ encode the data of the outer horizon (a Wilson loop along $\phi$ evaluates the entropy of the outer horizon), and it is unclear how to modify that computation to give the  ``size'' of an inner horizon since Wilson loops are independent of the radial position. This would be a very interesting puzzle to solve! 
\item {\it Interior points and distances:} The astute reader perhaps noticed that we only considered Wilson lines that start and end at the asymptotic boundaries, as depicted in Figure \ref{fig:lor1}. It is natural to ask why we never considered a Wilson line that terminates at some point in the interior. The answer is due to boundary conditions: the operator \eqref{eq:wilson1} is sensitive to the choice of initial and final state $\langle U_{i} |$. We only have an understanding of this choice once the endpoints reach the boundary, since there we can either explicitly test against the dual CFT (or at least motivate that certain symmetries should be preserved).   The very recent developments in \cite{Bhatta:2016hpz,Melnikov:2016eun} will be a useful first step in understanding more quantitatively the role of boundary conditions in $W_\sR(C_{ij})$.
\end{enumerate}

In addition, there are many other interesting future directions to explore which we discussed in section \ref{sec:appl}. 

\section*{Acknowledgements}
We would like to thank Per Kraus and Eric Perlmutter for many discussions on this topic, and Fotios Dimitrakopoulos for initial collaboration on this project. A.C. and E.L. are supported by Nederlandse Organisatie voor Wetenschappelijk Onderzoek (NWO) via a Vidi grant. We thank the Perimeter Institute for hospitality during the course of this project. This work is part of the Delta ITP consortium, a program of the NWO that is funded by the Dutch Ministry of Education, Culture and Science (OCW).


\appendix

\section{Conventions}\label{app:conv}

In general, we denote the 3 generators of $sl(2,\RR)$ as $\{L_0,L_{1},L_{-1}\}$. The algebra is given by
\be
[J_a,J_b]= \epsilon_{abc}J^c ~,\quad J^c=\delta^{cd}J_d~,
\ee
where $\epsilon_{abc}$ is a completely antisymmetric tensor and $\epsilon_{0+-}=1$; and  $\delta_{00}={1\over 2}$,  $\delta_{+-}=\delta_{-+}=-1$. In the fundamental representation, we use
\be
L_0= \left(\begin{array}{cc}1/2 & 0 \\ 0& -1/2\end{array}\right)~,\quad L_{1}= \left(\begin{array}{cc}0& 0 \\ -1& 0\end{array}\right)~,\quad L_{-1}= \left(\begin{array}{cc}0 & 1 \\ 0& 0\end{array}\right)~.
\ee

We label the $sl(3,\RR)$  generators as $T_a=\{L_i,W_m\}$ with $i=-1,0,1$ and $m=-2,\ldots,2$. The algebra reads
\bea
[L_i ,L_j] &=& (i-j)L_{i+j}~, \cr
[L_i, W_m] &=& (2i-m)W_{i+m}~, \cr
[W_m,W_n] &=& -{1 \over 3}(m-n)(2m^2+2n^2-mn-8)L_{m+n}~.
\eea
We work with the following matrices in the fundamental representation
\bea
L_1 & =& \left(\begin{array}{ccc} 0&0&0 \\ 1&0&0 \\ 0&1&0\end{array}\right),\quad L_0=  \left(\begin{array}{ccc}1&0&0 \\ 0&0&0 \\ 0&0&-1\end{array}\right),\quad  L_{-1} =  \left(\begin{array}{ccc} 0&-2&0 \\ 0&0&-2 \\ 0&0&0\end{array}\right)~,\cr &&\cr
W_2 &= &2  \left(\begin{array}{ccc} 0&0&0 \\ 0&0&0 \\ 1&0&0\end{array}\right),\quad W_1 = \left(\begin{array}{ccc} 0&0&0 \\ 1&0&0 \\ 0&-1&0 \end{array}\right),\quad  W_0 = {2\over 3}\left(\begin{array}{ccc} 1&0&0 \\ 0&-2&0 \\ 0&0&1\end{array}\right)~, \cr & &\cr
W_{-1} &= &\left(\begin{array}{ccc} 0&-2&0 \\ 0&0&2 \\ 0&0&0\end{array}\right),\quad W_{-2} = 2\left(\begin{array}{ccc} 0&0&4 \\ 0&0&0 \\ 0&0&0\end{array}\right)~. \label{sl3gens}
\eea
The quadratic traces are 
\bea
{\rm tr}_f ( L_0 L_0) &=& 2~,\quad   {\rm tr}_f ( L_1 L_{-1} ) = -4 ~,\cr   {\rm tr}_f ( W_0 W_0 )  &= &{8 \over 3}~,\quad {\rm tr}_f ( W_1 W_{-1} )  = -4 ~,\quad {\rm tr}_f ( W_2 W_{-2} )  = 16~.
\eea

The transition to the metric formulation of the theory is made via:
\be
 e = \ha\le(A - \bA\ri) ~,\qquad g_{\mu\nu} = \frac{1}{\Tr(L_0 L_0)} \Tr(e_{\mu} e_{\nu}) ~. \label{metconv}
\ee

The Wick rotation from Euclidean to Lorentzian time is
\be
w = f(r) e^{\frac{2\pi i}{\beta}\tau} \rightarrow -v~, \qquad \bw = f(r) e^{-\frac{2\pi i}{\beta}\tau} \rightarrow  u~ \label{app:wuv}
\ee
where $f(r)$ is an odd function that vanishes linearly at the black hole horizon and diverges at the AdS boundary. Following Figure \ref{fig:uv}, in  quadrant {\bf I} we have $u > 0$ and $v < 0$, which we parametrize in terms of a Lorentzian time coordinate $t_R$ as
\be
u = f(r) e^{\frac{2\pi}{\beta}t_R}~, \qquad v = - f(r)e^{-\frac{2\pi}{\beta} t_R} ~.\label{app:R1}
\ee
In quadrant {\bf IV} we have $u < 0$ and $v >0$, which we parametrize as
\be
u = -f(r) e^{-\frac{2\pi}{\beta}t_L}~, \qquad v =  f(r)e^{\frac{2\pi}{\beta} t_L} \label{app:L1} \ . 
\ee

\section{Thermofield states and KMS conditions}\label{app:kms}

In this appendix we review the definition  thermofield  state, and properties of thermal correlations functions. We will denote the relations discussed below as ``KMS conditions'' (even though only one of them is strictly speaking {\it the} KMS condition). 

Consider a system with a Hamiltonian $H$, and time-evolve operators in the Heisenberg picture:
\be
\sO(t) = e^{i H t} \sO(0) e^{-i H t}~.
\ee
It is very easy to show that for two operators $\sO_1$ and $\sO_2$, we have
\be
\Tr \le(e^{-\beta H} \sO_1(t - i \beta) \sO_2(0) \ri) = \Tr\le(e^{-\beta H} \sO_2(0) \sO_1(t)\ri)~. \label{unchargedKMS}
\ee
This is what one normally calls the KMS condition. 

Let us now try this for a different density matrix $\rho = e^{-\beta H - \beta \mu Q}$ with $Q$ another conserved charge of the system; for example, it could be a $U(1)$ charge, or $Q=W_0$ where $W_0$ is the zero mode of the ${\cal W}_3$ algebra. In this case we find
\begin{align}
\Tr\le(e^{-\beta H - \beta\mu Q} \sO_2(0) \sO_1(t) \ri) & = \Tr\le(e^{-\beta H - \beta\mu Q} \sO_2(0) e^{-\beta H - \beta \mu Q} e^{+ \beta H + \beta \mu Q} \sO_1(t) \ri) \cr
& = \Tr\le(\sO_2(0) e^{-\beta H - \beta \mu Q} \le[ e^{\beta \mu Q} \sO_1(t - i \beta) e^{-\beta \mu Q}\ri]\ri) \cr
& = \Tr\le(e^{-\beta H - \beta \mu Q} \sO_1(t - i \beta) \sO_2(0) \ri) e^{-\beta \mu q_1}~,
\end{align}
where in the last equality we have assumed that $\sO_1$ is an operator with definite charge $q_1$. Note that if $\sO_1$ was {\it not} a charge eigenstate we could stop at the line above and still get a useful (but more complicated) KMS relation. Thus the charged KMS relation is
\be
\Tr\le(e^{-\beta H - \beta \mu Q} \sO_1(t - i \beta) \sO_2(0) \ri) = \Tr\le(e^{-\beta H - \beta \mu Q} \sO_2(0) \sO_1(t)\ri) e^{+ \beta \mu q_1}~.
\ee
The extra factor involving the charge on the right-hand side appeared because of the mismatch between the Hamiltonian used to evolve the system (i.e. just $H$) and the Hamiltonian used to construct the density matrix (i.e. $H + \mu Q$). If we evolve the system using $H + \mu Q$ then there will be no extra factor involving the charge, and the correlator will be strictly periodic, as in \eqref{unchargedKMS}. 
 
The thermofield double state is defined as follows. Let ${\cal H}= {\cal H}_L\otimes {\cal H}_R$ denote the full Hilbert space which is composed by two copies of the original CFT Hilbert space. The thermofield state is defined by the following wave function on $\cal H$
\be
|\psi \rangle = \frac{1}{\sqrt{Z}}\sum_n e^{-\frac{\beta}{2}(E_n + \mu Q_n)} |{\sU} n\rangle_L \otimes |n \rangle_R~.
\ee
Here we included a chemical potential, and the sum is over all energy eigenstates of the system which carry as well $Q$ charge; $Z$ is a suitable normalization. $\sU$ is the anti-unitary operator that implements CPT; this is important since if one constructs the thermofield state by cutting open a path-integral then this CPT operator must be there (see e.g. \cite{Marolf:2013dba,Harlow:2014yka}). Anti-unitary implies that
\be
\sU^{-1} = \sU^{\dagger} \qquad \langle \sU \psi | \sU \phi \rangle = \langle \phi | \psi \rangle~, \label{antiU}
\ee
and the fact that $\sU$ implements CPT means
\be
\sU^{-1}\le(i H\ri)\sU = - i H \qquad \sU^{-1} \sO \sU \equiv \sO^{CPT} ~.\label{CPTness}
\ee
Note that $\sU$ actually commutes with $H$, but anticommutes with $i$. We denote the CPT conjugate of an operator with a superscript. For sake of simplicity, in the following we will consider scalar operators  and in this case $\sO^{CPT} =\sO^\dagger$.

Now let us carefully compute
\bea
\langle \psi| \sO_{1,L}(t_L) \sO_{2,R}(t_R) |\psi \rangle =  \frac{1}{Z}\sum_{m,n} \langle \sU n| e^{i \hat H t_L} \sO_{1} e^{-i \hat H t_L} |\sU m\rangle \langle n| e^{i \hat H t_R} \sO_{2} e^{-i \hat H t_R} |m \rangle e^{-\frac{\beta}{2}\le(\hat E_n + \hat E_m \ri)}~,
\eea
where $\hat E_n= E_n+\mu Q_n$. Note that we are evolving the system with $\hat H\equiv H+\mu Q$, which is the natural choice from the gravitational side. Looking at the first term, we find
\begin{align}
\langle \sU n| e^{i \hat H t_L} \sO_1 e^{-i \hat H t_L} |\sU m\rangle & = \langle \sU n| \sU e^{- i \hat H t_L} \sO_{1}^{CPT} e^{i \hat H t_L} m \rangle \cr
& = \langle e^{- i \hat H t_L} \sO_1^{CPT} e^{i \hat H t_L} m | n \rangle \cr
& = \langle m | e^{-i \hat H t_L} (\sO_1^{CPT})^{\dagger} e^{i \hat H t_L} |n \rangle\cr
& = \langle m | e^{-i \hat H t_L} \sO_1 e^{i \hat H t_L} |n \rangle~.
\end{align}
The first equality uses \eqref{CPTness} and the second uses \eqref{antiU}, the third equality  follows from the definition of the adjoint, and in the last line we used that  the operator is scalar. Thus we find
\begin{align}
\langle \psi| \sO_{1,L}(t_L) \sO_{2,R}(t_R) |\psi \rangle & = \sum_{m,n} e^{-\frac{\beta}{2}\le(\hat E_n + \hat E_m\ri) + i t_L(\hat E_n - \hat E_m) + i t_R(\hat E_n - \hat E_m)} \langle m | \sO_1 |n \rangle \langle n |\sO_2 |m \rangle \cr
& = \sum_{m,n} \langle m |e^{-\frac{\beta}{2}\le(\hat H - i \hat H(t_L + t_R)\ri)} \sO_1 e^{-\frac{\beta}{2}\le(\hat H + i \hat H (t_L + t_R)\ri)}|n \rangle \langle n | \sO_2 |m \rangle \cr
& = \Tr\le(e^{-\beta(H + \mu Q)} \sO_1\le(-t_L - \frac{i\beta}{2}\ri)  \sO_2(t_R)\ri)~. \label{twosidedthermal}
\end{align}
These manipulations shows how $\langle \psi| \sO_{1,L}(t_L) \sO_{2,R}(t_R) |\psi \rangle$ is related to the thermal correlation function. With some slight abuse of language, and in analogy to \eqref{unchargedKMS}, we will refer to this relation as a KMS condition. Note that the sign of $t_L$ is flipped: this relation explains what it means for ``time to run backwards on the other side''.

If instead we used $H$ to evolve the system, instead of $\hat H$, 
\be
\langle \psi| \sO_{1,L}(t_L) \sO_{2,R}(t_R) |\psi \rangle = \Tr\le(e^{-\beta H - \mu Q} \sO_1\le(-t_L - \frac{i\beta}{2}\ri) \sO_2(t_R)\ri) e^{-\frac{\beta \mu q_1}{2}}~, \label{kmstwosided}
\ee
where we assumed that $\sO_1$ is a scalar operator with a definite charge $q_1$.
 For operators with more complicated CPT conjugations or that are not charge eigenstates,  we would find more complicated versions of \eqref{twosidedthermal}.

From the above KMS conditions, we can derive further relation. Define the RR correlator as  a `one-sided' correlator in the thermofield state which   involves only operators on ${\cal H}_R$. For $\sO_1=\sO_2\equiv \sO$, we find
\be\label{CFTRR}
\langle \psi| \sO_{R}(t_f) \sO_{R}(t_i) |\psi \rangle = \Tr\left(e^{-{\beta}\hat H} \sO(t_f) \sO(t_i) \right)\,,
\ee
where we have suppressed the indexes $R$ in the right hand side of the equation since they are redundant. Analogously, the LL correlator is
\be\label{CFTLL}
\langle \psi| \sO_{L}(t_f) \sO_{L}(t_i) |\psi \rangle = \Tr\left(e^{-{\beta}\hat H} \sO(-t_f) \sO(-t_i) \right)\,.
\ee
For an LR correlator we have
\be\label{CFTLR}
\langle \psi| \sO_{L}(t_f) \sO_{R}(t_i) |\psi \rangle = \Tr\left(e^{-{\beta}\hat H} \sO(-t_f-i\beta/2) \sO(t_i) \right)\,,
\ee
and obviously, the RL correlator is given by
\be\label{CFTRL}
\langle \psi| \sO_{R}(t_f) \sO_{L}(t_i) |\psi \rangle = \Tr\left(e^{-{\beta}\hat H} \sO(t_f) \sO(-t_i-i\beta/2) \right)\,.
\ee
These previous identities imply that the correlators should be related as 
\bea
\langle \psi| \sO_{R}(t_f) \sO_{R}(t_i)|\psi \rangle &=& \langle \psi| \sO_{R}(t_f) \sO_{R}(t_i-i\beta)|\psi \rangle \cr &=&
\langle \psi| \sO_{L}(-t_f) \sO_{L}(-t_i) |\psi \rangle\cr &=&
\langle \psi| \sO_{L}(-t_f-i\beta/2) \sO_{R}(t_i) |\psi \rangle \cr &=& \langle\psi| \sO_{R}(t_f) \sO_{L}(-t_i-i\beta/2) |\psi \rangle\,,
\eea
The relations between the one-sided (RR and LL) and two-sided correlators (RL and LR) we denote as ``KMS conditions''. 


\section{Horizon gauge for ${\cal W}_3$ black hole}\label{app:gauge}

In this appendix we present the solution to the horizon condition constructed in \cite{Ammon:2011nk}.  This solution is valid for the non-rotating holomorphic black hole \eqref{hol}, however it is straight forward to check that it is also applicable for the non-rotating canonical black hole \eqref{can}. The ansatz used there is 
\begin{align}\label{BHgaugegb}
 A=g(r)^{-1}b(r)^{-1}(a_h+d)\,b(r)g(r)\,,\\ \nonumber
  \bar A=g(r)b(r)\,(\bar{a}_h+d)b(r)^{-1}g(r)^{-1}\,,
\end{align}
where $b(r)=e^{(r+r_0)L_0}$ with $e^{r_0}=\sqrt{2\pi{\cal L}/k}$, and they take
\begin{align}\label{BHgaugedef}
& g(r)=e^{F(r)(W_1-W_{-1})+G(r)L_0}\,,\\ \nonumber
& h(r)=e^{H(r)(W_1+W_{-1})}\,,
\end{align}
with $F(r)=F(-r),\,G(r)=G(-r),$ and $H(-r)=-H(-r)$; this implies that $g(r)=g(-r)$, $h(r)=h^{-1}(-r)$ and $h(0)=\mathds{1}$. Using \eqref{BHgaugedef}, a solution to \eqref{BHgauget}-\eqref{BHgaugex} is
\bea
Y^2&=&1+C\,\cosh^2(r)\,,\label{Y}\\
X&=&\sqrt{\frac{C-1+Y}{C-1-Y}}\,,\label{X}\\
G&=&-\frac{1}{Y}\log(X)\,\label{G}\\
\frac{F}{G}&=&\frac{\sqrt{C}}{2}\cosh(r)\,,\label{F}\\
\tan H&=&-\frac{\sinh(r)}{\sqrt{C-2-\cosh^2(r)}}\,.
\eea
 In this new radial parametrization, the asymptotic boundary is now located at $r=r_*$ which is given by
\be\label{newboundary}
\cosh^2(r_*)=C-2 \quad \longleftrightarrow \quad Y(r_*)=C-1\,.
\ee
In the BTZ limit, $C\rightarrow \infty$, we recover $r=r_*\rightarrow \infty$. 
From equations \eqref{Y}-\eqref{F}, we observe that the parameter $X$ diverges when $r=r_*$, and $Y$, $G$, and $F$ have an finite value.  At the boundary, we consider $X^{-1}$ as the cutoff $\varepsilon$, and we can express $Y$, $G$, and $F$ in terms of $C$. With these considerations, we diagonalize  $g(r_*)$, and find as eigenvalues:
\be
\lambda_g(r_*)=\left(\varepsilon^{-1}\,,\quad 1\,,\quad \varepsilon\right)=e^{-\log(\varepsilon)L_0}\,,
\ee
The eigenvectors of $g(r_*)$ are finite, i.e., they do not depend in $\varepsilon$. 
%


\section{Wilson line operator in AdS$_3$ higher spin gravity}\label{app:wilson}

This appendix is a brief summary of the results in \cite{Ammon:2013hba,Castro:2014mza} with emphasize on how to evaluate the Wilson line. To recap, the operator is defined as 
\be\label{Wilsondef}
W_\sR(y_i,y_j)= \langle U_i | {\cal P} \exp\le( \int_{C_{ij}} A\ri){\cal P} \exp\le(\int_{C_{ij}} \bar A\ri) | U_f \rangle ~.
\ee
$\sR$ is an infinite dimensional representation of the gauge group, and $C_{ij}$ is a curve with bulk endpoints $(y_i,y_j)$.  $U(y)$ is a probe field that lives in the worldline $C_{ij}$, and which quantum numbers are governed by $\sR$. Its boundary values are chosen such that $U_i=U_f=\mathds{1}$: this choice ensures that the Wilson line induces a conical deficit in the background and the answer is Lorentz invariant. In a saddle point approximation, the value of the Wilson line is
\be\label{Wilsonalpha}
-\log W_{\mathcal{R}}(C)=\Tr(\log(M) P_0)~,
\ee
where $P_0$ is the conjugated momentum of the probe field $U$. More importantly $P_0$ carries the data related to the Casimir's of the representation $\sR$: for example in $sl(N)\times sl(N)$ a highest weight representation labelled by  quantum numbers $(h, w_s)=(\bar h,\bar w_s)$ we would have
\be
P_0 = \frac{h}{2} L_0 + \sum_{s=3}^{N} \frac{w_s}{2}W^{(s)}_0~.
\ee
%
%
Here $W^{(s)}_0$ are the Cartan elements of $sl(N,\R)$; $h$ is the conformal dimension of the probe $U$ and $w_s$ corresponds to a higher spin charge. The matrix $M$ in \eqref{Wilsonalpha} contains the information about the background connections $(A,\bar A)$:
\be
M \equiv R(y_i)L(y_i)L^{-1}(y_f)R^{-1}(y_f)~,
\ee 
with $R(y)$ and $L(y)$ defined according to \eqref{RL0}. This expression makes evident that the Wilson line is only sensitive to the endpoints of $C_{ij}$.

%
%

We will restrict now the discussion to Wilson lines in $sl(3)\times sl(3)$. As we send the endpoints of the Wilson line to one of the two boundaries, located at $r\rightarrow\pm\infty$, we only need to consider the asymptotic behavior of the eigenvalues of $M$ to evaluate \eqref{Wilsonalpha}. If asymptotically we have
\be\label{bex}
b(r)=\bar b(r)\underset{r\to \infty} \to e^{r L_0}~,
\ee
the eigenvalues of $M$ will asymptote to
\begin{align}\label{m1m2.1}
\lambda_M\sim\left(m_1\, \epsilon^{-4}\,,\quad \frac{m_2}{m_1} \,,\quad \frac{ \epsilon^{4}}{m_2}\right)\,,
\end{align}
where $\epsilon=e^{-\rho}$ is the cutoff, and $m_1$ and $m_2$ are related to the coefficients of the characteristic polynomial as:
 \be\label{m1m2.2}
c_1=\text{Tr}_f(M)=m_1\, \epsilon^{-4}+...\,,\qquad c_2=\frac{1}{2}\left(\text{Tr}_f(M)^2-\text{Tr}_f(M^2)\right)=m_2\,\epsilon^{-4}+...\,\,.
 \ee
Note that $m_1=m_1(y_i,y_f)$ and $m_2=m_2(y_i,y_f)$ depend on the endpoints and the background charges carried by the connections. The asymptotic behaviour of the Wilson line close to the boundary is given by
\be\label{Wilsonm1m2}
-\log W_\sR(y_i,y_f)= \frac{h}{2}\,\log\left(\frac{m_1m_2(y_i,y_f)}{\epsilon^8}\right) + {w_3}\,\log\left(\frac{m_1(y_i,y_f)}{m_2(y_i,y_f)}\right)\,.
\ee
where we kept only universal terms as $\epsilon\to 0$. 

It is interesting to note that for  $\Delta \phi=0$,  the solutions depends only $a_t$ and $\bar a_t$, which are elements of $SL(2,\mathbb{R})$  due to the holonomy condition \eqref{smoothness}.  Therefore,  $M$ belongs as well to $SL(2,\mathbb{R})$ which implies that $m_1=m_2$ and 
\be\label{Wilsonm1m2t}
-\log W_\sR(t_i,t_f)= {h}\,\log\left(\frac{m_1(t_i,t_f)}{\epsilon^4}\right)\,.
\ee
%
 %
%

In general we only need that at infinity
\be\label{bassymp}
b(r),\,\bar b(r) \cong e^{ -\log(\varepsilon) L_0}\,,
\ee
where $\cong$ means equal up to conjugation, and $\varepsilon$ controls the UV cutoff as we approach the asymptotic boundaries.  If the conjugation matrices do not depend on $\varepsilon$, the formulas \eqref{m1m2.1}-\eqref{Wilsonm1m2t} hold with the substitution $\epsilon$ by $\varepsilon$. This is the case of the black hole gauge, detailed in Section \ref{app:gauge}.


\section{Computation of Kruskal gauge for higher spin black hole} \label{app:kruskal}
Here we provide details of the computation of the radial functions $b(\rho)$ and $\bb(\rho)$ that are required to put the higher spin black hole in Kruskal gauge. The basic constraint on these functions arises from the demand that the Euclidean objects defined as 
\be
B(r,\tau) = e^{a_{\tau} \tau} b(r) e^{-i L_0 \frac{2\pi \tau}{\beta}} \qquad \bB(r, \tau) = e^{i L_0 \frac{2 \pi \tau}{\beta}}\bb(\rho)e^{- \ba_{\tau}\tau} \label{Bexp}
\ee\
be smooth functions of the complex coordinates
\be
w = r e^{\frac{2\pi i \tau}{\beta}} \qquad \bw = r e^{-\frac{2\pi i\tau}{\beta}} \label{chol}
\ee
near the Euclidean origin. By smooth, we mean that the expansion of $B(w,\bw)$ contains only positive integer powers of $w, \bw$. As described in the bulk text, this analyticity property guarantees that the gauge connections can be analytically continued to a Lorentzian section that describes a two-sided black hole with a smooth horizon. 

On the other hand, to have a clean CFT interpretation of the bulk connections, we need to also demand that as it approaches the boundary $b(r)$ blow up as $b(r) \sim \exp(g(r) L_0)$ with $g(r)$ some function that tends to infinity at the boundary. Here we describe the construction of the functions $b, \bb$ that satisfy these two requirements. 

\subsection{Setup}

First, we use coordinates where the horizon is at $r = 0$, and which further match onto the more conventional $\rho$ coordinate at large $r$ as $\rho = e^{r}$. In other words the function $g(r) = e^{r}$. Now consider diagonalizing $a$ and $\ba$: the holonomy condition tells us that $a_{\tau}$ and $\ba_{\tau}$ are conjugate to $L_0$, so we have
\be
a_{\tau} = V \le(\frac{2 \pi i L_0}{\beta} \ri)V^{-1} \qquad \ba_{\tau} = \bV \le(\frac{2 \pi i L_0}{\beta}\ri) \bV^{-1} \ . 
\ee
Inserting these expansions into \eqref{Bexp} we find
\be
B(r,\tau) = V e^{\frac{2\pi i \tau}{\beta}L_0} V^{-1} b(r) e^{-i\frac{2\pi \tau}{\beta}L_0},
\ee
and a similar expression for $\bB$. We will focus for now on $B$. We expand
\be
V^{-1} b(r) = \exp\le(\sum_{a} F_a(r) T^a\ri) \label{Fdef}
\ee
where $a$ runs over the generators of the algebra, $F_a(r)$ is a set of mode functions to be defined shortly, and the $T^a$ are the generators. Note that the demand that $B$ depend smoothly on $w, \bw$ as defined in \eqref{chol} ties together the time and radial dependence. In this basis the time-dependence is simply a conjugation by $L_0$, multiplying each generator by a factor of $e^{-\frac{2 \pi i h(a) \tau}{\beta}}$, where $h(a)$ is the weight under $L_0$ of the generator $T^a$, i.e. $[L_0, T^{a}] = -h(a) T^a$. Thus the analyticity condition requires that near the origin we have:
\be
F_a(r \to 0) \sim r^{|h(a)|} \ . 
\ee
as well as a parity condition on $r$ (i.e. $F_a(r)$ should be either even or odd). 

We also require that at infinity we approach $b(r \to \infty) \sim \exp\le(e^{r} L_0\ri)$. It is convenient to define a basis of functions $f_a^m(r)$ such that
\be
f_a^m(r \to 0) \sim  r^{|h(a)|} \qquad f_a^m(r \to \infty) \sim e^{-m r} \ . 
\ee
Such a basis is presented explicitly below and is easy to find as the functions are otherwise unconstrained. We now further expand
\be
F_a = \sum_{m = -1,0,\cdots} c_{a}^m f_a^m(r)
\ee
By adjusting the coefficients $c_a^m$ we may reproduce any function at infinity to a prescribed order in an expansion in inverse powers in $e^{-r}$. We will calculate only the terms $m = -1,0$ as this is sufficient to calculate any correlator in $SL(3)$ higher spin gravity: for $SL(N)$ we require $N-1$ terms. 

\subsection{Diagonalization}
We now explicitly calculate the matrix logarithm of 
\be
Q \equiv V^{-1} \exp(\rho L_0)
\ee
to the first two orders in inverse powers of $\rho \equiv e^{r}$ to find the expansion coefficients $c_a^m$. 

It is easiest to diagonalize $Q$ and take the logarithm of the eigenvalues. To diagonalize $Q$ in the asymptotic limit we follow an algorithm somewhat similar to that normally used in quantum mechanical perturbation theory, with some modifications arising from the fact that $Q$ is not Hermitian. Define $x \equiv e^{\rho}$. Denoting the $i$-th eigenvalue and eigenvector as $\lam^{(i)}$ and $v^{(i)}$ respectively, we expand everything in powers of $x$ to find:
\be
\le(Q_1 x + Q_0 + Q_{-1} x^{-1}\cdots\ri) \le( v^{(i)}_0 + v^{(i)}_{-1} x^{-1} + \cdots\ri) = (\lam_1^{(i)}x + \lam_0^{(i)} + \cdots)\le( v^{(i)} + v^{(i)}_{-1} x^{-1} + \cdots\ri),
\ee
The $Q_\al$ may be found explicitly and directly diagonalized without much difficulty. The challenge is to extract from the $Q_\al$ the behavior of the $v^{(i)}$. We assume the expansion in powers of the eigenvectors starts at $\sO(x^0)$: this can always be arranged by rescaling the individual eigenvectors. We will determine each $v^{(i)}$ only to leading order, i.e. $v^{(i)}_0$. 

We first need to first determine the scaling behavior of the eigenvalues.  Note first that if we define $\mathcal{Q}_n \equiv \Tr(Q^n)$, then the characteristic polynomial of $Q$ is given by
\be
P_Q(\lam) = -\lam^3 + \mathcal{Q}_1 \lam^2 - \ha (\mathcal{Q}_1^2 - \mathcal{Q}_2)\lam + 1,
\ee
where we have used the fact that product of the eigenvalues is $1$ as $Q \in SL(3)$. From direct computation we now find that as $x \to \infty$, $\mathcal{Q}_1 \sim q_1 x + \sO(1)$, $\ha(\mathcal{Q}_1^2 - \mathcal{Q}_2) \sim q_2 x + \sO(1)$, where $q_1$ and $q_2$ are presented explicitly below. Now by balancing terms in the characteristic polynomial we find that the eigenvalues scale as
\be
\lam^{(i)}(x \to \infty) \sim  \le(q_1 x,\frac{q_2}{q_1},\frac{1}{q_2 x}\ri),
\ee
Thus we see that the order of the starting term in $\lam^{(i)}_\al$ depends on which eigenvalue we are studying, e.g. $\lam^{(1)}_1 = q_1$ but $\lam^{(2)}_1 = 0$. 

We turn now to the eigenvectors. We begin with the largest eigenvalue, $\lam^{(1)}$. The $\sO(x^1)$ equation is
\be
Q_1 v^{(1)}_0 = \lam_1^{(1)} v^{(1)}_0
\ee 
In other words, $v^{(1)}_0$ is an eigenvector of $Q_1$ itself with eigenvalue $\lam_1^{(1)} = q_1$. If we now examine the eigenvectors of $Q_1$ we see that it has two eigenvectors with zero eigenvalue, $\mbox{Dim}(\mbox{Ker}(Q_1)) = 2$, as well as a single nonzero eigenvector with eigenvalue $q_1$: thus $v^{(1)}_0$ is fixed to be this eigenvector. 

We turn now to $\lam^{(2)}$. The $\sO(x^1)$ equation is now
\be
Q_1 v_0^{(2)} = 0 
\ee 
and thus we find only that $v_0^{(2)}$ belongs to the two-dimensional kernel of $Q_1$ discussed above. To narrow it down within this subspace, we study the $\sO(x^0)$ equation, which is
\be
Q_1 v_{-1}^{(2)} + \le(Q_0 - \lam_0^{(2)} {\bf 1}\ri) v_{0}^{(2)} = 0
\ee
Thus $\le(Q_0 - \lam_0^{(2)} {\bf 1}\ri) v_{0}^{(2)}$ lies within the image of $Q_1$: but this means that it is proportional to the only eigenvector of $Q_1$ with nonzero eigenvalue, and so is proportional to $v^{(1)}_0$ found above. So we see that
\be
v_{0}^{(2)} = \le(Q_0 - \lam_0^{(2)} {\bf 1}\ri)^{-1} v^{(1)}_0
\ee
fixing it up to rescaling. We turn finally to the eigenvector corresponding to $\lam^{(3)}$. While presumably the above procedure can be systematized to arbitrarily higher order, as $\lam^{(3)}$ is the last eigenvalue, we may use a trick: the $\sO(x^1)$ and $\sO(x^0)$ equations are
\be
Q_1 v^{(3)}_0 = 0 \qquad Q_2 v^{(3)}_0 = 0 \ . 
\ee
and thus $v^{(3)}_0$ lies in the intersection of the kernel of $Q_1$ and the kernel of $Q_2$; we may explicitly check that this intersection is a one-dimensional subspace, fixing $v^{(3)}_0$.

Thus we have determined the eigenvalues and eigenvectors. We write now
\be
Q(x \to \infty) \sim W\exp(D)W^{-1} = \exp(W D W^{-1})
\ee
where $W$ is the matrix whose columns are the $v^{(i)}$ and $D$ is the following diagonal matrix:
\be
D = \log(\mbox{diag}(\lam^{(i)})) =  L_0 (\log x)+ \Lambda \qquad \Lambda \equiv \mbox{diag}\le(\log q_1,\log\le(\frac{q_2}{q_1}\ri),-\log(q_2)\ri) 
\label{diagEig} \  
\ee
We now equate $W D W^{-1}$ with $\sum_{a} F_a(r) T^a$ defined in $\eqref{Fdef}$, multiply by $T^b$, and take a trace to find
\be
\sum_{a} \sum_{m = -1,0}\delta^{ab}c^m_a f^m_a(r \to \infty) = \Tr(W D W^{-1} T^b) \
\ee
where we have defined the Killing metric on the Lie algebra as $\delta^{ab} \equiv \Tr(T^a T^b)$ and its inverse by $\delta_{ab}$. Now from the explicit form of the $W$'s and of the mode functions $f^m_a$ we can find the expansion coefficients $c^m_a$. Note that due to the judicious choice of the mode functions, the term $m = -1$ corresponds directly to the $L_0 (\log x)$ term in \eqref{diagEig} and the $m =0$ term to the second (constant in $x$) term. 
\be
c_{a}^{-1} = \delta_{ab} \Tr(W L_0 W^{-1} T^b) \qquad c_a^{0} = \delta_{ab} \Tr(W \Lambda W^{-1} T^b) \label{cform}
\ee
From \eqref{Fdef} this fixes $b(r)$ as 
\be
b(r) = V \exp\le(\sum_a \sum_{m = -1,0} c_{a}^m f_a^m(r)\ri)
\ee

For the barred connection we follow precisely the same procedure to find instead
\be
\bar{b}(r) = \exp\le(\sum_a \sum_{m = -1,0} \bar{c}_{a}^m f_a^m(r)\ri)\bV^{-1} \ . 
\ee

\subsection{Details}
Here we present (some of) the results from implementing the algorithm above. The mode functions that we use are
\be
f^1_{L_0} = f^1_{W_0} = 2 \cosh(r) \qquad f^1_{L_\pm 1} = f^1_{W_{\pm 1}} = 2 \sinh(r) \qquad f^1_{W_{\pm 2}} = 4\le(\sinh^2\le(\frac{r}{2}\ri) + \ha \tanh^2\le(\frac{r}{2}\ri)\ri)
\ee
Note that each $f^1_a(r \to \infty) \sim e^{r} + \sO(e^{-r})$. We have also
\be
f^0_{L_0} = f^0_{W_0} = 1 \qquad f^0_{L\pm 1} = f^0_{W_{\pm 1}} = \tanh(r) \qquad f^{0}_{W_{\pm 2}} = \tanh^2(r),
\ee
so that $f^0_{a}(r \to \infty) \sim 1 + \sO(e^{-2r}).$ 

The equations that follow are lengthy and rather unenlightening. Note that $V$, $\bV$ are only defined up to rescaling of their individual columns (subject to the constraint that they each have unit determinant). Some attempts were made to use this freedom to reduce the complexity of the ensuing algebra. It is likely that a solution with less complexity exists, but we did not make a serious attempt to find it. 

\subsubsection{Unbarred sector} 
We take $V$ to be:
\be
V = \left(
\begin{array}{ccc}
 \frac{\left(C+\sqrt{4 C-3}-1\right) \sqrt{\sL} \sqrt{2 \pi }}{\sqrt[3]{C-3} \sqrt{C} \sqrt[6]{4 C-3}} & \frac{\sqrt{2} (C-2) \sqrt{\sL} \sqrt{\pi }}{\sqrt[3]{C-3} \sqrt{C} \sqrt[6]{4 C-3}} & \frac{\left(C-\sqrt{4 C-3}-1\right) \sqrt{\sL} \sqrt{2 \pi }}{\sqrt[3]{C-3} \sqrt{C} \sqrt[6]{4 C-3}} \\
 \frac{-\sqrt{4 C-3}-1}{2 \sqrt[3]{C-3} \sqrt[6]{4 C-3}} & -\frac{1}{\sqrt[3]{C-3} \sqrt[6]{4 C-3}} & \frac{\sqrt{4 C-3}-1}{2 \sqrt[3]{C-3} \sqrt[6]{4 C-3}} \\
 \frac{\sqrt{C}}{2 \sqrt[3]{C-3} \sqrt[6]{4 C-3} \sqrt{\sL} \sqrt{2 \pi }} & -\frac{\sqrt{C}}{2 \sqrt[3]{C-3} \sqrt[6]{4 C-3} \sqrt{\sL} \sqrt{2 \pi }} & \frac{\sqrt{C}}{2 \sqrt[3]{C-3} \sqrt[6]{4 C-3} \sqrt{\sL} \sqrt{2 \pi }} \\
\end{array}
\right)
\ee
Now in computing $V^{-1} \exp(\rho \sL_0)$ we find 
\be
q_1 = \frac{\sqrt{C} \left(\sqrt{2} \sqrt{4 C-3}-3 \sqrt{2}\right)}{8 \sqrt{\pi } (C-3)^{2/3} \sqrt[3]{4 C-3} \sqrt{\sL}} \qquad q_2 = \frac{\sqrt{C}}{2 \sqrt{2 \pi } \sqrt[3]{C-3} \sqrt[6]{4 C-3} \sqrt{\sL}}
 \ee 
The matrix $W$ of eigenvectors is then found to be:
\be
\left(
\begin{array}{ccc}
 0 & 0 & \frac{\sqrt{4 C-3}-3}{\sqrt{4 C-3}+3} \\
 0 & \frac{(4 C-3)^{5/6}-C \sqrt[3]{4 C-3}}{\sqrt[3]{C-3} (C-1)} & \frac{1}{\frac{1}{2}+\frac{3}{2 \sqrt{4 C-3}}} \\
 1 & \frac{(4 C-3)^{5/6}-C \sqrt[3]{4 C-3}}{\sqrt[3]{C-3} (C-1)} & 1 \\
\end{array}
\right)
\ee

\subsubsection{Barred sector} 
We take $\bV$ to be:
\be
\left(
\begin{array}{ccc}
 \frac{C^{5/2} ((C-6) C+4) \sqrt{\sL}}{\sqrt[3]{\frac{(C-3) C^3 \sqrt{4 C-3}}{C-2}} \left(C+\sqrt{4 C-3}-1\right) \sqrt[6]{2 \pi }} & \frac{C}{(C-2) \sqrt[3]{\frac{(C-3) C^3 \sqrt{4 C-3}}{C-2}} \sL (2 \pi )^{2/3}} & -\frac{C \sqrt[3]{\pi }}{2^{2/3} \sqrt[3]{\frac{(C-3) C^3 \sqrt{4 C-3}}{C-2}} \left(\pi  \sqrt{4 C-3} \sL-C \pi  \sL+\pi  \sL\right)} \\
 \frac{C^2 \left(-\sqrt{4 C-3} C+3 C+2 \sqrt{4 C-3}-2\right) \sL \sqrt[3]{\pi }}{2^{2/3} \sqrt[3]{\frac{(C-3) C^3 \sqrt{4 C-3}}{C-2}}} & \frac{\sqrt{C}}{\sqrt[6]{2} (C-2) \sqrt[3]{\frac{(C-3) C^3 \sqrt{4 C-3}}{C-2}} \sqrt{\sL} \sqrt[6]{\pi }} & \frac{\sqrt{C} \left(C \left(\sqrt{4 C-3}+3\right)-2 \left(\sqrt{4 C-3}+1\right)\right)}{2 \sqrt[3]{\frac{(C-3) C^3 \sqrt{4 C-3}}{C-2}} ((C-6) C+4) \sqrt{\sL} \sqrt[6]{2 \pi }} \\
 \frac{C^{3/2} ((C-6) C+4) \sL^{3/2} \pi ^{5/6}}{\sqrt[6]{2} \sqrt[3]{\frac{(C-3) C^3 \sqrt{4 C-3}}{C-2}}} & -\frac{\sqrt[3]{\pi }}{2^{2/3} \sqrt[3]{\frac{(C-3) C^3 \sqrt{4 C-3}}{C-2}}} & \frac{\sqrt[3]{\pi }}{2^{2/3} \sqrt[3]{\frac{(C-3) C^3 \sqrt{4 C-3}}{C-2}}} \\
\end{array}
\right)
\ee
In computing $\exp(\rho \sL_0) \bV$ we find
\be
\bar{q}_1 = \frac{\sqrt[3]{C-2} C^{3/2} \left(C-\sqrt{4 C-3}-1\right) \sqrt{\sL}}{\sqrt[6]{2 \pi } \sqrt[3]{C-3} \sqrt[6]{4 C-3}} \qquad \bar{q}_2 = \frac{C \left(\sqrt{4 C-3} C-C-4 \sqrt{4 C-3}\right)}{2 \sqrt[3]{2 \pi } (C-3)^{2/3} \sqrt[3]{C-2} \sqrt[3]{4 C-3}}
\ee
The matrix of eigenvectors is
\be
\bar{W} = \left(
\begin{array}{ccc}
 \frac{\sqrt{4 C-3} C+C-4 \sqrt{4 C-3}}{2 (C-2) C^{3/2} \sqrt{4 C-3} ((C-6) C+4) \sL^{3/2} \sqrt{2 \pi }} & -\frac{\sqrt{C} \sqrt[6]{4 C-3} \left(\sqrt{4 C-3}-3\right) \sqrt{\sL} \sqrt[6]{\pi }}{2^{5/6} \left(\frac{C-3}{C-2}\right)^{2/3}} & 1 \\
 -1 & \frac{2 \sqrt[3]{C-3} (C-2)^{5/3} C^2 \sqrt[6]{4 C-3} \left(-C+\sqrt{4 C-3}+1\right)^2 \sL^2 (2 \pi )^{2/3}}{C \left(\sqrt{4 C-3}-1\right)-4 \sqrt{4 C-3}} & 0 \\
 \frac{C \left(\frac{1}{\sqrt{4 C-3}}-1\right)+4}{4-2 C} & 0 & 0 \\
\end{array}
\right)
\ee

From here it is straightforward to use \eqref{cform} (and a computer) to find the expansion coefficients $c_a^m, \bar{c}_a^m$. The results are, however, too lengthy to write down. 
\bibliographystyle{utphys}
\bibliography{HigherSpin}

\providecommand{\href}[2]{#2}\begingroup\raggedright\begin{thebibliography}{10}

\bibitem{Gutperle:2011kf}
M.~Gutperle and P.~Kraus, ``{Higher Spin Black Holes},''
  \href{http://dx.doi.org/10.1007/JHEP05(2011)022}{{\em JHEP} {\bfseries 1105}
  (2011) 022},
\href{http://arxiv.org/abs/1103.4304}{{\ttfamily arXiv:1103.4304 [hep-th]}}.

\bibitem{Ammon:2013hba}
M.~Ammon, A.~Castro, and N.~Iqbal, ``{Wilson Lines and Entanglement Entropy in
  Higher Spin Gravity},'' \href{http://dx.doi.org/10.1007/JHEP10(2013)110}{{\em
  JHEP} {\bfseries 1310} (2013) 110},
\href{http://arxiv.org/abs/1306.4338}{{\ttfamily arXiv:1306.4338 [hep-th]}}.

\bibitem{deBoer:2013vca}
J.~de~Boer and J.~I. Jottar, ``{Entanglement Entropy and Higher Spin Holography
  in AdS$_3$},'' \href{http://dx.doi.org/10.1007/JHEP04(2014)089}{{\em JHEP}
  {\bfseries 04} (2014) 089},
\href{http://arxiv.org/abs/1306.4347}{{\ttfamily arXiv:1306.4347 [hep-th]}}.

\bibitem{deBoer:2014sna}
J.~de~Boer, A.~Castro, E.~Hijano, J.~I. Jottar, and P.~Kraus, ``{Higher spin
  entanglement and $ {\mathcal{W}}_{\mathrm{N}} $ conformal blocks},''
  \href{http://dx.doi.org/10.1007/JHEP07(2015)168}{{\em JHEP} {\bfseries 07}
  (2015) 168},
\href{http://arxiv.org/abs/1412.7520}{{\ttfamily arXiv:1412.7520 [hep-th]}}.

\bibitem{Banados:2016nkb}
M.~Banados, R.~Canto, and S.~Theisen, ``{Higher Spin Black Holes in Three
  Dimensions: Comments on Asymptotics and Regularity},''
\href{http://arxiv.org/abs/1601.05827}{{\ttfamily arXiv:1601.05827 [hep-th]}}.

\bibitem{Engelhardt:2015fwa}
D.~Engelhardt, B.~Freivogel, and N.~Iqbal, ``{Electric fields and quantum
  wormholes},'' \href{http://dx.doi.org/10.1103/PhysRevD.92.064050}{{\em Phys.
  Rev.} {\bfseries D92} no.~6, (2015) 064050},
\href{http://arxiv.org/abs/1504.06336}{{\ttfamily arXiv:1504.06336 [hep-th]}}.

\bibitem{Harlow:2015lma}
D.~Harlow, ``{Wormholes, Emergent Gauge Fields, and the Weak Gravity
  Conjecture},'' \href{http://dx.doi.org/10.1007/JHEP01(2016)122}{{\em JHEP}
  {\bfseries 01} (2016) 122},
\href{http://arxiv.org/abs/1510.07911}{{\ttfamily arXiv:1510.07911 [hep-th]}}.

\bibitem{Guica:2015zpf}
M.~Guica and D.~L. Jafferis, ``{On the construction of charged operators inside
  an eternal black hole},''
\href{http://arxiv.org/abs/1511.05627}{{\ttfamily arXiv:1511.05627 [hep-th]}}.

\bibitem{Ammon:2011nk}
M.~Ammon, M.~Gutperle, P.~Kraus, and E.~Perlmutter, ``{Spacetime Geometry in
  Higher Spin Gravity},'' \href{http://dx.doi.org/10.1007/JHEP10(2011)053}{{\em
  JHEP} {\bfseries 1110} (2011) 053},
\href{http://arxiv.org/abs/1106.4788}{{\ttfamily arXiv:1106.4788 [hep-th]}}.

\bibitem{Ammon:2012wc}
M.~Ammon, M.~Gutperle, P.~Kraus, and E.~Perlmutter, ``{Black holes in three
  dimensional higher spin gravity: A review},''
  \href{http://dx.doi.org/10.1088/1751-8113/46/21/214001}{{\em J.Phys.}
  {\bfseries A46} (2013) 214001},
\href{http://arxiv.org/abs/1208.5182}{{\ttfamily arXiv:1208.5182 [hep-th]}}.

\bibitem{deBoer:2013gz}
J.~de~Boer and J.~I. Jottar, ``{Thermodynamics of higher spin black holes in
  $AdS_3$},'' \href{http://dx.doi.org/10.1007/JHEP01(2014)023}{{\em JHEP}
  {\bfseries 1401} (2014) 023},
\href{http://arxiv.org/abs/1302.0816}{{\ttfamily arXiv:1302.0816 [hep-th]}}.

\bibitem{Bunster:2014mua}
C.~Bunster, M.~Henneaux, A.~Perez, D.~Tempo, and R.~Troncoso, ``{Generalized
  Black Holes in Three-dimensional Spacetime},''
\href{http://arxiv.org/abs/1404.3305}{{\ttfamily arXiv:1404.3305 [hep-th]}}.

\bibitem{deBoer:2014fra}
J.~de~Boer and J.~I. Jottar, ``{Boundary Conditions and Partition Functions in
  Higher Spin AdS$_3$/CFT$_2$},''
\href{http://arxiv.org/abs/1407.3844}{{\ttfamily arXiv:1407.3844 [hep-th]}}.

\bibitem{deBoer:1998ip}
J.~de~Boer, ``{Six-dimensional supergravity on S**3 x AdS(3) and 2-D conformal
  field theory},'' \href{http://dx.doi.org/10.1016/S0550-3213(99)00160-1}{{\em
  Nucl.Phys.} {\bfseries B548} (1999) 139--166},
\href{http://arxiv.org/abs/hep-th/9806104}{{\ttfamily arXiv:hep-th/9806104
  [hep-th]}}.

\bibitem{Henneaux:2010xg}
M.~Henneaux and S.-J. Rey, ``{Nonlinear $W_{infinity}$ as Asymptotic Symmetry
  of Three-Dimensional Higher Spin Anti-de Sitter Gravity},''
  \href{http://dx.doi.org/10.1007/JHEP12(2010)007}{{\em JHEP} {\bfseries 1012}
  (2010) 007},
\href{http://arxiv.org/abs/1008.4579}{{\ttfamily arXiv:1008.4579 [hep-th]}}.

\bibitem{Campoleoni:2010zq}
A.~Campoleoni, S.~Fredenhagen, S.~Pfenninger, and S.~Theisen, ``{Asymptotic
  symmetries of three-dimensional gravity coupled to higher-spin fields},''
  \href{http://dx.doi.org/10.1007/JHEP11(2010)007}{{\em JHEP} {\bfseries 1011}
  (2010) 007},
\href{http://arxiv.org/abs/1008.4744}{{\ttfamily arXiv:1008.4744 [hep-th]}}.

\bibitem{Gaberdiel:2011wb}
M.~R. Gaberdiel and T.~Hartman, ``{Symmetries of Holographic Minimal Models},''
  \href{http://dx.doi.org/10.1007/JHEP05(2011)031}{{\em JHEP} {\bfseries 1105}
  (2011) 031},
\href{http://arxiv.org/abs/1101.2910}{{\ttfamily arXiv:1101.2910 [hep-th]}}.

\bibitem{Campoleoni:2011hg}
A.~Campoleoni, S.~Fredenhagen, and S.~Pfenninger, ``{Asymptotic W-symmetries in
  three-dimensional higher-spin gauge theories},''
  \href{http://dx.doi.org/10.1007/JHEP09(2011)113}{{\em JHEP} {\bfseries 1109}
  (2011) 113},
\href{http://arxiv.org/abs/1107.0290}{{\ttfamily arXiv:1107.0290 [hep-th]}}.

\bibitem{Castro:2011fm}
A.~Castro, E.~Hijano, A.~Lepage-Jutier, and A.~Maloney, ``{Black Holes and
  Singularity Resolution in Higher Spin Gravity},''
  \href{http://dx.doi.org/10.1007/JHEP01(2012)031}{{\em JHEP} {\bfseries 1201}
  (2012) 031},
\href{http://arxiv.org/abs/1110.4117}{{\ttfamily arXiv:1110.4117 [hep-th]}}.

\bibitem{Hijano:2014sqa}
E.~Hijano and P.~Kraus, ``{A new spin on entanglement entropy},''
  \href{http://dx.doi.org/10.1007/JHEP12(2014)041}{{\em JHEP} {\bfseries 12}
  (2014) 041},
\href{http://arxiv.org/abs/1406.1804}{{\ttfamily arXiv:1406.1804 [hep-th]}}.

\bibitem{Banados:2012ue}
M.~Banados, R.~Canto, and S.~Theisen, ``{The Action for higher spin black holes
  in three dimensions},'' \href{http://dx.doi.org/10.1007/JHEP07(2012)147}{{\em
  JHEP} {\bfseries 1207} (2012) 147},
\href{http://arxiv.org/abs/1204.5105}{{\ttfamily arXiv:1204.5105 [hep-th]}}.

\bibitem{Perez:2012cf}
A.~Perez, D.~Tempo, and R.~Troncoso, ``{Higher spin gravity in 3D: Black holes,
  global charges and thermodynamics},''
  \href{http://dx.doi.org/10.1016/j.physletb.2013.08.038}{{\em Phys.Lett.}
  {\bfseries B726} (2013) 444--449},
\href{http://arxiv.org/abs/1207.2844}{{\ttfamily arXiv:1207.2844 [hep-th]}}.

\bibitem{Compere:2013nba}
G.~Comp{\`e}re, J.~I. Jottar, and W.~Song, ``{Observables and Microscopic
  Entropy of Higher Spin Black Holes},''
  \href{http://dx.doi.org/10.1007/JHEP11(2013)054}{{\em JHEP} {\bfseries 1311}
  (2013) 054},
\href{http://arxiv.org/abs/1308.2175}{{\ttfamily arXiv:1308.2175 [hep-th]}}.

\bibitem{Henneaux:2013dra}
M.~Henneaux, A.~Perez, D.~Tempo, and R.~Troncoso, ``{Chemical potentials in
  three-dimensional higher spin anti-de Sitter gravity},''
  \href{http://dx.doi.org/10.1007/JHEP12(2013)048}{{\em JHEP} {\bfseries 1312}
  (2013) 048},
\href{http://arxiv.org/abs/1309.4362}{{\ttfamily arXiv:1309.4362 [hep-th]}}.

\bibitem{Banados:2015tft}
M.~Banados, A.~Castro, A.~Faraggi, and J.~I. Jottar, ``{Extremal Higher Spin
  Black Holes},''
\href{http://arxiv.org/abs/1512.00073}{{\ttfamily arXiv:1512.00073 [hep-th]}}.

\bibitem{Kraus:2011ds}
P.~Kraus and E.~Perlmutter, ``{Partition functions of higher spin black holes
  and their CFT duals},'' \href{http://dx.doi.org/10.1007/JHEP11(2011)061}{{\em
  JHEP} {\bfseries 1111} (2011) 061},
\href{http://arxiv.org/abs/1108.2567}{{\ttfamily arXiv:1108.2567 [hep-th]}}.

\bibitem{Gaberdiel:2012yb}
M.~R. Gaberdiel, T.~Hartman, and K.~Jin, ``{Higher Spin Black Holes from
  CFT},'' \href{http://dx.doi.org/10.1007/JHEP04(2012)103}{{\em JHEP}
  {\bfseries 1204} (2012) 103},
\href{http://arxiv.org/abs/1203.0015}{{\ttfamily arXiv:1203.0015 [hep-th]}}.

\bibitem{David:2012iu}
J.~R. David, M.~Ferlaino, and S.~P. Kumar, ``{Thermodynamics of higher spin
  black holes in 3D},'' \href{http://dx.doi.org/10.1007/JHEP11(2012)135}{{\em
  JHEP} {\bfseries 1211} (2012) 135},
\href{http://arxiv.org/abs/1210.0284}{{\ttfamily arXiv:1210.0284 [hep-th]}}.

\bibitem{Israel:1976ur}
W.~Israel, ``{Thermo field dynamics of black holes},''
\href{http://dx.doi.org/10.1016/0375-9601(76)90178-X}{{\em Phys. Lett.}
  {\bfseries A57} (1976) 107--110}.

\bibitem{Maldacena:2001kr}
J.~M. Maldacena, ``{Eternal black holes in anti-de Sitter},'' {\em JHEP}
  {\bfseries 0304} (2003) 021,
\href{http://arxiv.org/abs/hep-th/0106112}{{\ttfamily arXiv:hep-th/0106112
  [hep-th]}}.

\bibitem{Kraus:2012uf}
P.~Kraus and E.~Perlmutter, ``{Probing higher spin black holes},''
  \href{http://dx.doi.org/10.1007/JHEP02(2013)096}{{\em JHEP} {\bfseries 1302}
  (2013) 096},
\href{http://arxiv.org/abs/1209.4937}{{\ttfamily arXiv:1209.4937 [hep-th]}}.

\bibitem{Hegde:2015dqh}
A.~Hegde, P.~Kraus, and E.~Perlmutter, ``{General Results for Higher Spin
  Wilson Lines and Entanglement in Vasiliev Theory},''
  \href{http://dx.doi.org/10.1007/JHEP01(2016)176}{{\em JHEP} {\bfseries 01}
  (2016) 176},
\href{http://arxiv.org/abs/1511.05555}{{\ttfamily arXiv:1511.05555 [hep-th]}}.

\bibitem{Chamblin:1999tk}
A.~Chamblin, R.~Emparan, C.~V. Johnson, and R.~C. Myers, ``{Charged AdS black
  holes and catastrophic holography},''
  \href{http://dx.doi.org/10.1103/PhysRevD.60.064018}{{\em Phys. Rev.}
  {\bfseries D60} (1999) 064018},
\href{http://arxiv.org/abs/hep-th/9902170}{{\ttfamily arXiv:hep-th/9902170
  [hep-th]}}.

\bibitem{Hartnoll:2009sz}
S.~A. Hartnoll, ``{Lectures on holographic methods for condensed matter
  physics},'' \href{http://dx.doi.org/10.1088/0264-9381/26/22/224002}{{\em
  Class. Quant. Grav.} {\bfseries 26} (2009) 224002},
\href{http://arxiv.org/abs/0903.3246}{{\ttfamily arXiv:0903.3246 [hep-th]}}.

\bibitem{Castro:2015csg}
A.~Castro, D.~M. Hofman, and N.~Iqbal, ``{Entanglement Entropy in Warped
  Conformal Field Theories},''
  \href{http://dx.doi.org/10.1007/JHEP02(2016)033}{{\em JHEP} {\bfseries 02}
  (2016) 033},
\href{http://arxiv.org/abs/1511.00707}{{\ttfamily arXiv:1511.00707 [hep-th]}}.

\bibitem{Castro:2014tta}
A.~Castro, S.~Detournay, N.~Iqbal, and E.~Perlmutter, ``{Holographic
  entanglement entropy and gravitational anomalies},''
  \href{http://dx.doi.org/10.1007/JHEP07(2014)114}{{\em JHEP} {\bfseries 07}
  (2014) 114},
\href{http://arxiv.org/abs/1405.2792}{{\ttfamily arXiv:1405.2792 [hep-th]}}.

\bibitem{Banados:1992gq}
M.~Banados, M.~Henneaux, C.~Teitelboim, and J.~Zanelli, ``{Geometry of the
  (2+1) black hole},'' \href{http://dx.doi.org/10.1103/PhysRevD.48.1506}{{\em
  Phys.Rev.} {\bfseries D48} (1993) 1506--1525},
\href{http://arxiv.org/abs/gr-qc/9302012}{{\ttfamily arXiv:gr-qc/9302012
  [gr-qc]}}.

\bibitem{Kraus:2002iv}
P.~Kraus, H.~Ooguri, and S.~Shenker, ``{Inside the horizon with AdS / CFT},''
  \href{http://dx.doi.org/10.1103/PhysRevD.67.124022}{{\em Phys. Rev.}
  {\bfseries D67} (2003) 124022},
\href{http://arxiv.org/abs/hep-th/0212277}{{\ttfamily arXiv:hep-th/0212277
  [hep-th]}}.

\bibitem{Datta:2014ska}
S.~Datta, J.~R. David, M.~Ferlaino, and S.~P. Kumar, ``{Higher spin
  entanglement entropy from CFT},''
\href{http://arxiv.org/abs/1402.0007}{{\ttfamily arXiv:1402.0007 [hep-th]}}.

\bibitem{Callan:2012ip}
R.~Callan, J.-Y. He, and M.~Headrick, ``{Strong subadditivity and the covariant
  holographic entanglement entropy formula},''
  \href{http://dx.doi.org/10.1007/JHEP06(2012)081}{{\em JHEP} {\bfseries 1206}
  (2012) 081},
\href{http://arxiv.org/abs/1204.2309}{{\ttfamily arXiv:1204.2309 [hep-th]}}.

\bibitem{Knuttel:Thesis2014}
J.~Knuttel, ``Entanglement entropy and black holes in (2+1)-dimensional higher
  spin gravity,'' Master's thesis, University of Amsterdam, The Netherlands,
  2014.

\bibitem{Castro:2014mza}
A.~Castro and E.~Llabr{\'e}s, ``{Unravelling Holographic Entanglement Entropy
  in Higher Spin Theories},''
  \href{http://dx.doi.org/10.1007/JHEP03(2015)124}{{\em JHEP} {\bfseries 1503}
  (2015) 124},
\href{http://arxiv.org/abs/1410.2870}{{\ttfamily arXiv:1410.2870 [hep-th]}}.

\bibitem{Gaberdiel:2013jca}
M.~R. Gaberdiel, K.~Jin, and E.~Perlmutter, ``{Probing higher spin black holes
  from CFT},'' \href{http://dx.doi.org/10.1007/JHEP10(2013)045}{{\em JHEP}
  {\bfseries 1310} (2013) 045},
\href{http://arxiv.org/abs/1307.2221}{{\ttfamily arXiv:1307.2221 [hep-th]}}.

\bibitem{Hartman:2013qma}
T.~Hartman and J.~Maldacena, ``{Time Evolution of Entanglement Entropy from
  Black Hole Interiors},''
  \href{http://dx.doi.org/10.1007/JHEP05(2013)014}{{\em JHEP} {\bfseries 05}
  (2013) 014},
\href{http://arxiv.org/abs/1303.1080}{{\ttfamily arXiv:1303.1080 [hep-th]}}.

\bibitem{Calabrese:2005in}
P.~Calabrese and J.~L. Cardy, ``{Evolution of entanglement entropy in
  one-dimensional systems},''
  \href{http://dx.doi.org/10.1088/1742-5468/2005/04/P04010}{{\em J. Stat.
  Mech.} {\bfseries 0504} (2005) P04010},
\href{http://arxiv.org/abs/cond-mat/0503393}{{\ttfamily arXiv:cond-mat/0503393
  [cond-mat]}}.

\bibitem{Liu:2013iza}
H.~Liu and S.~J. Suh, ``{Entanglement Tsunami: Universal Scaling in Holographic
  Thermalization},''
  \href{http://dx.doi.org/10.1103/PhysRevLett.112.011601}{{\em Phys. Rev.
  Lett.} {\bfseries 112} (2014) 011601},
\href{http://arxiv.org/abs/1305.7244}{{\ttfamily arXiv:1305.7244 [hep-th]}}.

\bibitem{Liu:2013qca}
H.~Liu and S.~J. Suh, ``{Entanglement growth during thermalization in
  holographic systems},''
  \href{http://dx.doi.org/10.1103/PhysRevD.89.066012}{{\em Phys. Rev.}
  {\bfseries D89} no.~6, (2014) 066012},
\href{http://arxiv.org/abs/1311.1200}{{\ttfamily arXiv:1311.1200 [hep-th]}}.

\bibitem{Roberts:2014ifa}
D.~A. Roberts and D.~Stanford, ``{Two-dimensional conformal field theory and
  the butterfly effect},''
  \href{http://dx.doi.org/10.1103/PhysRevLett.115.131603}{{\em Phys. Rev.
  Lett.} {\bfseries 115} no.~13, (2015) 131603},
\href{http://arxiv.org/abs/1412.5123}{{\ttfamily arXiv:1412.5123 [hep-th]}}.

\bibitem{Maldacena:2015waa}
J.~Maldacena, S.~H. Shenker, and D.~Stanford, ``{A bound on chaos},''
\href{http://arxiv.org/abs/1503.01409}{{\ttfamily arXiv:1503.01409 [hep-th]}}.

\bibitem{Perlmutter:2016pkf}
E.~Perlmutter, ``{Bounding the Space of Holographic CFTs with Chaos},''
\href{http://arxiv.org/abs/1602.08272}{{\ttfamily arXiv:1602.08272 [hep-th]}}.

\bibitem{Iqbal:2011in}
N.~Iqbal, H.~Liu, and M.~Mezei, ``{Semi-local quantum liquids},''
  \href{http://dx.doi.org/10.1007/JHEP04(2012)086}{{\em JHEP} {\bfseries 04}
  (2012) 086},
\href{http://arxiv.org/abs/1105.4621}{{\ttfamily arXiv:1105.4621 [hep-th]}}.

\bibitem{Faulkner:2011tm}
T.~Faulkner, N.~Iqbal, H.~Liu, J.~McGreevy, and D.~Vegh, ``{Holographic
  non-Fermi liquid fixed points},''
  \href{http://dx.doi.org/10.1098/rsta.2010.0354}{{\em Phil. Trans. Roy. Soc.}
  {\bfseries A 369} (2011) 1640},
\href{http://arxiv.org/abs/1101.0597}{{\ttfamily arXiv:1101.0597 [hep-th]}}.

\bibitem{Faulkner:2009wj}
T.~Faulkner, H.~Liu, J.~McGreevy, and D.~Vegh, ``{Emergent quantum criticality,
  Fermi surfaces, and AdS(2)},''
  \href{http://dx.doi.org/10.1103/PhysRevD.83.125002}{{\em Phys. Rev.}
  {\bfseries D83} (2011) 125002},
\href{http://arxiv.org/abs/0907.2694}{{\ttfamily arXiv:0907.2694 [hep-th]}}.

\bibitem{Bhatta:2016hpz}
A.~Bhatta, P.~Raman, and N.~V. Suryanarayana, ``{Holographic Conformal Partial
  Waves as Gravitational Open Wilson Networks},''
\href{http://arxiv.org/abs/1602.02962}{{\ttfamily arXiv:1602.02962 [hep-th]}}.

\bibitem{Melnikov:2016eun}
D.~Melnikov, A.~Mironov, and A.~Morozov, ``{On skew tau-functions in higher
  spin theory},''
\href{http://arxiv.org/abs/1602.06233}{{\ttfamily arXiv:1602.06233 [hep-th]}}.

\bibitem{Marolf:2013dba}
D.~Marolf and J.~Polchinski, ``{Gauge/Gravity Duality and the Black Hole
  Interior},'' \href{http://dx.doi.org/10.1103/PhysRevLett.111.171301}{{\em
  Phys. Rev. Lett.} {\bfseries 111} (2013) 171301},
\href{http://arxiv.org/abs/1307.4706}{{\ttfamily arXiv:1307.4706 [hep-th]}}.

\bibitem{Harlow:2014yka}
D.~Harlow, ``{Jerusalem Lectures on Black Holes and Quantum Information},''
\href{http://arxiv.org/abs/1409.1231}{{\ttfamily arXiv:1409.1231 [hep-th]}}.

\end{thebibliography}\endgroup

\end{document}